\documentclass[aps,prl,twocolumn]{revtex4-1}
\usepackage{amssymb}
\usepackage{epsfig}
\usepackage{graphicx}
\usepackage{amsmath}
\usepackage{amsfonts}

\begin{document}




\title{Vibrations, Quanta and Biology}

%
\author{S.~F. Huelga and M.~B. Plenio$^{a,b}$ $^{\ast}$
\\\vspace{6pt}
$^{a}${\em{Institut f{\"u}r Theoretische Physik, Universit{\"a}t Ulm, Albert-Einstein-Allee 11, 89073 Ulm, Germany}}
$^{b}${\em{Center for Integrated Quantum Science and Technologies, Albert-Einstein-Allee 11, 89073 Ulm, Germany}}\\\vspace{6pt}\received{v3.0 released January 2010} }


\begin{abstract}
\vspace*{2ex}
Quantum biology is an emerging field of research that concerns itself with the experimental and theoretical exploration of non-trivial quantum phenomena in biological systems. In this tutorial overview we aim to bring out fundamental assumptions and questions in the field, identify basic design principles and develop a key underlying theme -- the dynamics of quantum dynamical networks in the presence of an environment and the fruitful interplay that the two may enter. At the hand of three biological phenomena whose understanding is held to require quantum mechanical processes, namely excitation and charge transfer in photosynthetic complexes, magneto-reception in birds and the olfactory sense, we demonstrate that this underlying theme encompasses them all, thus suggesting its wider relevance as an archetypical framework for quantum biology.\\

\bigskip
\bigskip

\end{abstract}

\maketitle

\section{Introduction}\label{Intro}
%
Following early speculations concerning the potential role of quantum physics in biology \cite{Jordan43}, recent progress in science and technology has led to the rapid emergence of a new direction of research whose aim is the experimental and theoretical exploration of quantum effects in biology (see e.g. \cite{MohseniOEP13,Solvay11}) which are taking place on length and timescales that allow quantum dynamics and environmental fluctuations to enter an intricate and fruitful interplay.

Before we enter into more detailed discussions, let us first make some points as to why the existence of quantum effects in biology may be considered surprising and why there is rapidly growing excitement for developing what is being called {\em quantum biology}.

To begin with, biological systems are, almost by definition, open systems, as they need to be continuously supplied with energy to maintain the out of equilibrium state that life represents. Open systems, however, especially warm, wet and noisy biological systems, are subject to environmental fluctuations that are usually expected to result in fast decoherence and, as a result, the suppression of well controlled quantum dynamics. Thus quantum phenomena may at first sight seem unlikely to play a significant role in biology. There are arguments however to counter this pessimistic view. At the level of molecular complexes and proteins, processes that are of fundamental importance for biological function can be very fast (taking place within picoseconds) and well localised (extending across a few nanometers, the size of proteins) and may therefore exhibit quantum phenomena before the environment has had an opportunity to destroy them. Furthermore, early work in quantum information science, for example, has shown that thermal noise in stationary non-equilibrium systems may in fact support the existence of quantum coherence and entanglement \cite{PlenioH02,HartmannDB06}. Hence the possible existence of significant quantum dynamics is not only a question of sufficiently short length and time scales but may also depend on a constructive interplay between a quantum dynamical system and its environment such that quantum correlations are not simply washed out or suppressed but may in fact be enhanced or regenerated by the interaction with the environment.

These arguments suggest that quantum effects in biology are possible at the right length- and time scales. Indeed, quantum phenomena such as electron tunneling \cite{DeVaultC66,DeVault80} have been observed in biological systems and there is some evidence for proton tunneling in enzymes (see e.g. \cite{ChaMK89}). As such, tunneling phenomena are not intimately related with biology. Electron tunneling for example is a well-known and important phenomenon in solid state physics. The question thus remains whether, on the one hand, biological systems will exhibit more complex quantum dynamical phenomena that may either involve several interacting particles or multiple interacting components of a network or, on the other hand, whether the specifics of the biological systems and their environments will play a crucial role in allowing or supporting certain quantum dynamical phenomena in biology. Only then would we call these "non-trivial" quantum effects in biological systems.

We do expect however that in the course of evolution Nature will have learnt to make use of quantum phenomena only if these enable or make more efficient a useful biological function
that provides an evolutionary advantage. It is indeed well established from a quantum information perspective that pure quantum dynamics of multi-component systems can provide qualitative performance improvements over classical systems for example where transport is concerned \cite{Kempe03}. This provides further support for the expectation that nature has developed non-trivial quantum phenomena in the dynamics of biological systems, possibly supported by their environment whose presence and influence is unavoidable. These quantum phenomena are not merely a by-product of the underlying quantum nature of chemical bonds but are actually exploited by biological systems to enhance performance and achieve novel functionalities. The clear demonstration that Nature makes use of quantum effects would bring about the necessity for a significant change of thinking for biologists as they would be required to grasp quantum concepts in order to understand some fundamental biological processes. The very same fact would however also present the opportunity to learn from biology by unraveling the mechanisms by which quantum dynamics and its interplay with environments lead to enhanced performance. The resulting design principles have the potential to lead to the development of new applications at the bio-nano scale \cite{ScholesFO+11}.

\begin{figure}[hbt]
\centerline{\includegraphics[width=10cm]{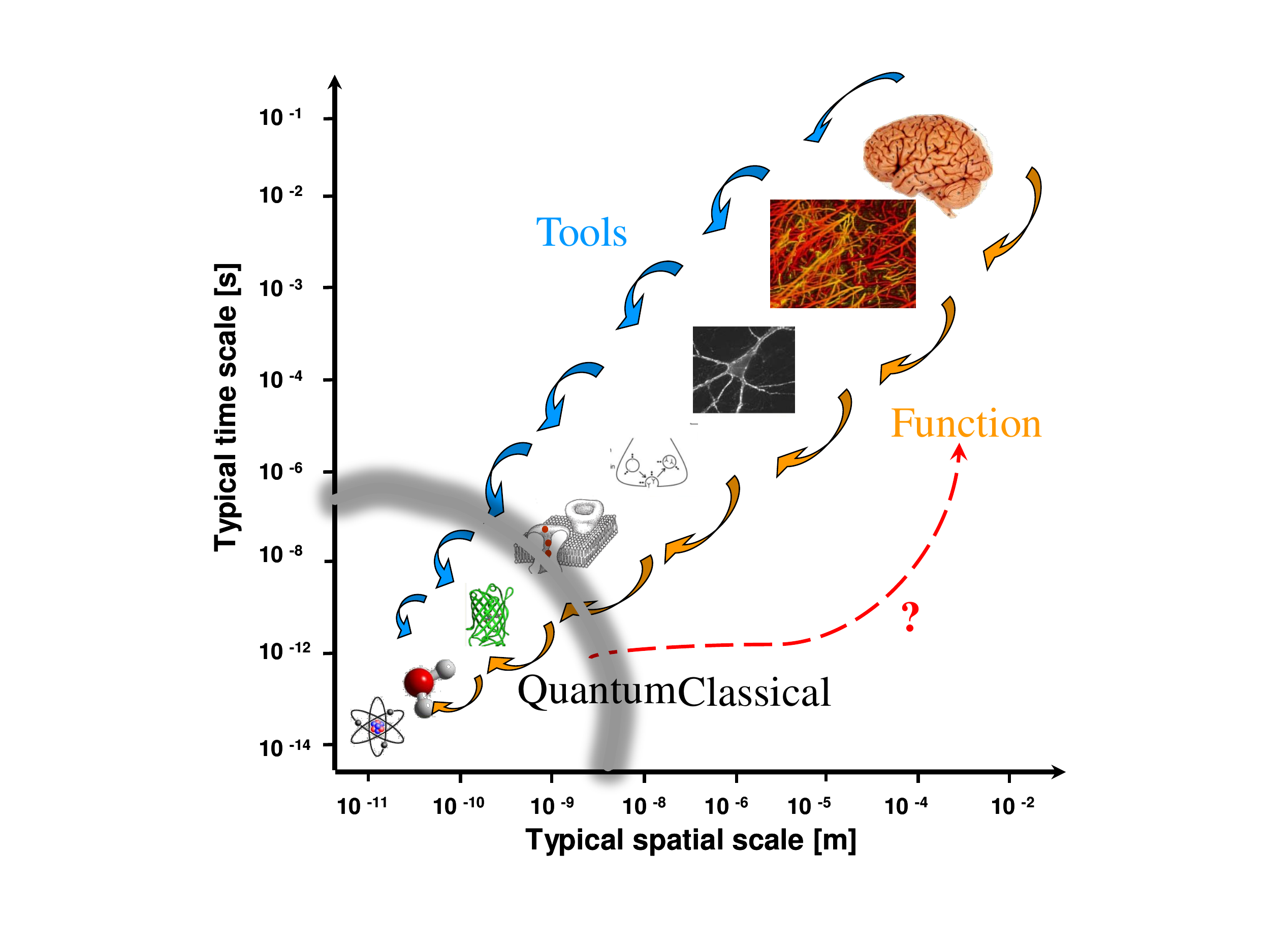}}
\vspace*{-0.75cm}
\caption{Biological systems are organized in hierarchical structures. The continuous refinement of experimental tools permits the investigation of ever finer detail giving rise to the discovery of novel phenomena. At a certain level we expect quantum physical properties to become relevant. Whether nature has evolved to enhance them to take benefit from them (quantum enhanced efficiency) or to suppress them to avoid their detrimental effects (quantum noise) represents one of interesting open question at the heart of quantum biology. (Figure courtesy of Alipasha Vaziri.)}
\label{landfig}
\end{figure}

Last but not least, the recent acceleration of the development of quantum biology also has a very practical, technological reason. Indeed, it is worthwhile noting that quantum biology is not a new field but does go back a long time, perhaps to Jordan's book "Die Physik und das Geheimnis des Lebens" \cite{Jordan43} in which he posed the question "Sind die Gesetze der Atomphysik und Quantenphysik f{\"u}r die Lebensvorg{\"a}nge von wesentlicher Bedeutung?" (Are the laws of atomic and quantum physics of essential importance for life?) and coined the term Quanten-Biologie (quantum biology). So, why this renewed and rapidly growing interest in the field? To understand this, it is helpful to remember the development of quantum information science, whose theoretical foundations had been studied for some time  by the 1990's. This research had revealed that concepts of quantum information have the potential to provide real performance advantages over classical systems. Crucially, however, it also became clear that quantum technologies had advanced sufficiently to turn these theoretical ideas into reality. This led to the emergence of a rapid development of both theory and experiment which is continuing to this day.

In recent years, quantum biology too has been benefitting considerably from the refinement of experimental tools that are beginning to provide direct access to the observation of quantum dynamics in biological systems \cite{EngelCR+07,MercerEK+09,PanitchayangkoonHF+10,ColliniWW+10} thanks to their increasing sensitivity to quantum phenomena at short length and time scales (see fig. \ref{landfig} for a suggestive illustration of this point). These newly found technological capabilities have helped to elevate the study of quantum biology from a largely theoretical endeavor to a field in which theoretical questions, concepts and hypotheses may be tested experimentally and thus be subjected to experimental verification or falsification. Indeed we should stress here that experiments are essential (even more so than in the well-controlled present day systems of quantum information science) to verify theoretical models because biological systems under investigation have a complexity and structural variety that prevents us from knowing and controlling all their aspects in detail. Results obtained by these refined experimental techniques lead to new theoretical challenges and thus stimulate the development of novel theoretical approaches. It is this mutually beneficial interplay between experiment and theory that promises an accelerated development of the field.

In this tutorial overview we will explore the type of phenomena that currently define the field of quantum biology and aim to bring out what are key questions at the present stage of development. We will then focus our discussions on a principle that is rapidly gaining recognition as being of central importance in this field -- the crucial role of the interplay between quantum dynamics of multi-site systems, a.k.a. networks, on the one hand and of complex, structured environments on the other. We will illustrate the generality of this principle by using it to analyse and interpret three key examples of quantum effects in biological systems, environment assisted excitation energy transport in photosynthesis \cite{EngelCR+07,MohseniRL+08,PlenioH08}, magneto-reception of birds \cite{SchultenSW78,RitzAS00,RitzTP+04} and the mechanism underlying olfaction \cite{Turin96,FrancoTM+11}, for which there is some compelling theoretical and/or experimental evidence that quantum phenomena are essential for their understanding. These three examples will hopefully stimulate the discovery of many more biological phenomena for which non-trivial quantum effects are of fundamental importance and thus come to be seen as the seeds from which a rich phenomenology of quantum effects in biology may grow.
%
\begin{figure}[hbt]
\centerline{\includegraphics[width=8cm]{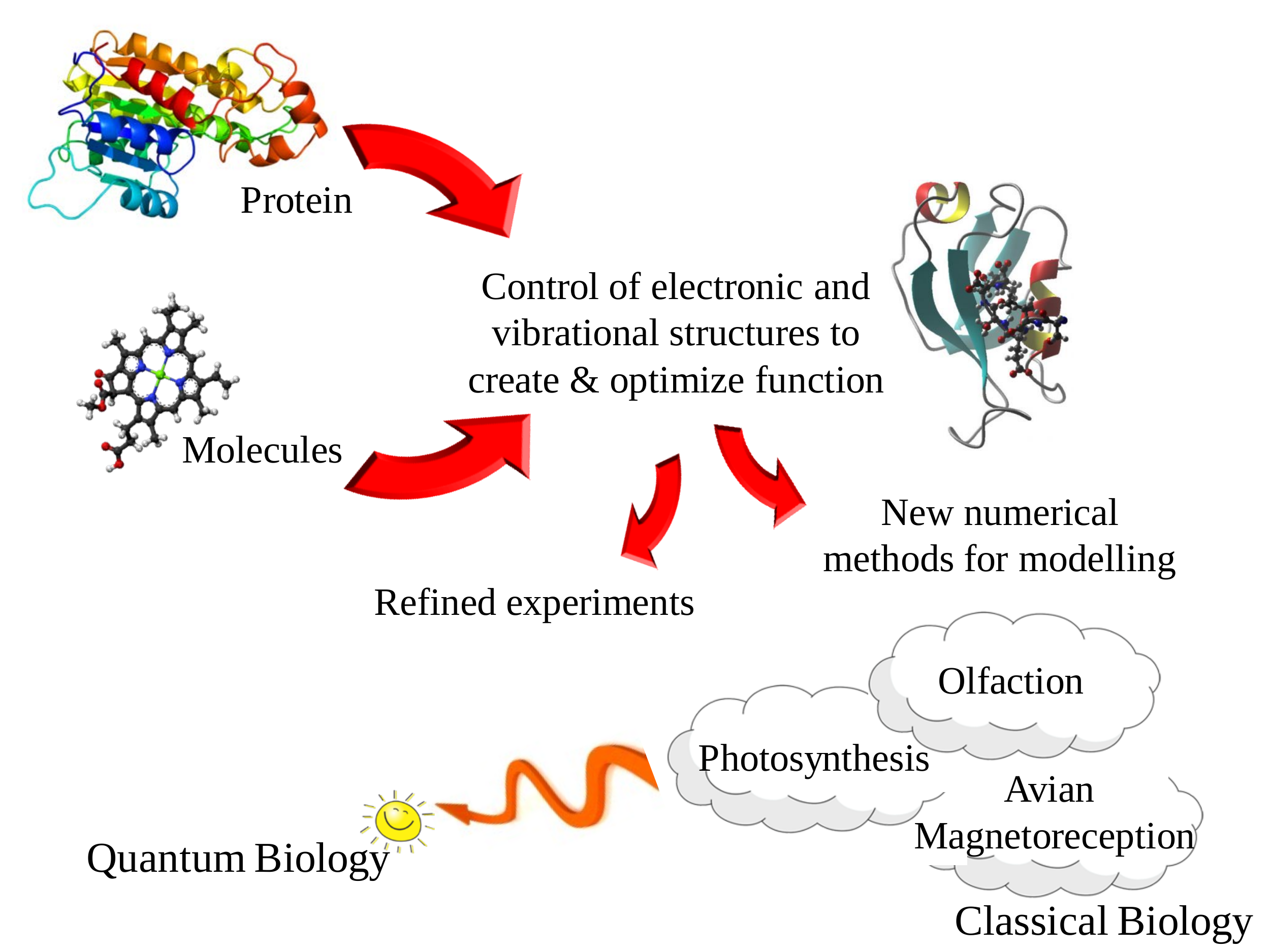}}
\caption{Cartoon illustrating the broad line of argument in this article: Following the identification of general questions, we will argue that biological systems use protein structure to adjust the properties of transport or sensory networks and, at the same time, those of the environment of these networks. Mutual tuning of these structures through evolutionary adaptation may achieve optimal performance which in turn can be explained from generalizable design principles. These design principles can lead to the formulation of novel structures and experiments to amplify and verify quantum effects. The accurate description of the interplay of structured environments and quantum dynamics especially in the non-perturbative regime requires the development of novel theoretical methods. These concepts will be discussed and shown to apply to the current three examples of quantum effects in biology, photosynthesis, avian magneto-reception and olfaction. It is the hope that these examples will be joined by many others and lead to the emergence of a new research branch, quantum biology.}
\label{Opt}
\end{figure}

\section{General Questions and Principles}
In the following we would like to draw attention to a number of general points related to the study of quantum effects in biology and identify broad questions that might be worth exploring.

{\em Biology, dynamics, transport ---} Although perhaps obvious, we would like to stress that biology is not merely about static structures but that the dynamics that is enabled by these highly organized structures plays a key role. Indeed, this dynamics is essential as biological systems need to be supplied with energy and drained of entropy to maintain the out-of-equilibrium state that life represents. This requires a wide variety of transport processes including excitation energy transfer but extending to charge transport including electrons, protons and ions as well as the transport of larger molecules, peptides and proteins. But the role of transport processes extends much further as they are also essential ingredients in processes such as signal recognition and transduction which requires the transfer of excitations or real physical particles. Hence transport and more general dynamical phenomena play a fundamental role in biology and therefore in the following discussion of quantum effects in biology.

{\em Quantum traits \& advantages ---} Quantum physics offers a wide variety of features that distinguish it from its classical counterpart and which may in some cases allow for enhanced performance and functionalities. This includes coherence and interference, that is, wave-like features of particles which may lead to faster propagation for example in quantum random walks \cite{Kempe03}. More sophisticated multi-particle coherence phenomena distributed across different components of a system such as entanglement hold the potential to achieve higher sensitivity to external signals \cite{WinelandBI+92,HuelgaMP+97}. Another quantum trait, the quantization of energy, leads to well-defined energy levels that can be excited in discrete portions, quanta of energy, only. In the exchange of energy between two quantum systems these features will then enforce highest efficiency if the respective energy levels, such as vibrational eigenfrequencies, are well matched. This in turn permits the unambiguous identification of frequencies and thus can facilitate the construction of sensing devices \cite{Turin96}.

Therefore elucidating to what extent and under what conditions quantum traits may be realised in biological systems and how they are exploited for enhanced performance is of considerable interest.

{\em Open quantum systems ---} Biological structures in general are not isolated as they are open systems that are in permanent contact with their environments. While many combinations of system and environment are conceivable, in this text the system will tend to be formed of electronic degrees of freedom (excitons, electrons, ...) while the environment will either be of vibrational nature or composed of electron and nuclear spin degrees of freedom. While thermal fluctuations imparted by the environment may be beneficial for example to overcome potential barriers between distinct classical configurations and therefore facilitating processes such as protein folding or chemical reactions, the benefit of system-environment interactions is less evident in the quantum world. The uncontrolled fluctuations of these environments will lead to changes in the local structures in which the electronic degrees of freedom are embedded and may therefore lead to decoherence. The advantages of quantum coherence in random walks \cite{Kempe03} may for example be destroyed by decoherence \cite{KeatingLMW06} and therefore the normal expectation in quantum technologies tended to be that almost complete isolation of the quantum system from its environment is required to reap the benefits of quantum effects. However, it was realized in that field that the interaction between system and environment may lead to the creation of quantum properties such as coherence and entanglement \cite{PlenioH02} and investigations in quantum biology found that they may enhance transport for example in photosynthetic quantum networks \cite{MohseniRL+08,PlenioH08}.

The elucidation of mechanisms by which the quantum dynamics on networks may enter a fruitful interplay with their environment to achieve enhanced performance and long-lived quantum coherence in transport, signalling and sensing is thus of fundamental importance.

{\em Environments are structured ---} As we shall discover, the role of the environment in biological systems is not a passive one. Biological environments are not featureless sources of white noise nor do they represent merely weak perturbations. The environmental spectral densities \cite{Footnote0} describing their interaction with the system tend to display two principal structures, a broad smooth background which has a short memory time and interacts with the system mainly through its fluctuations, i.e. causing noise, and well defined narrow features corresponding, for example, to long-lived vibrational motion that can lead to quasi-coherent dynamical non-equilibrium exchanges between system and environment. As we will see later the smooth background is due mostly to the protein environment as well as noise processes originating from solvents while the sharp features tend to originate from long-lived vibrations that belong to molecules that are held within the protein scaffold \cite{CaycedoSolerCA+12,RatsepCR+11}.

The rich structure of biological environments leads to non-Markovian dynamics which is also non-perturbative being neither weak nor strong compared to the intra-system dynamics. These features deviate from environments that are typically studied in quantum technologies and uncovering non-trivial consequences of these structures that may have been exploited by nature and are therefore worthy of careful study.

{\em Out of equilibrium \& back-action ---} Many processes of interest to quantum biology are initiated by the sudden generation of an initial excitation, such as a quantum of energy to be transported, which drives the system away from equilibrium. As a consequence of this sudden excitation neither the system nor its environment will remain stationary. The ensuing out-of-equilibrium dynamics then leads to perturbation of the environment by the system and a back-action of the very same environment on the system initiating as a result a dynamical exchange between system and environment. If the perturbation of the environment concerns the broad smooth background it will relax back very rapidly to its equilibrium value due to its short memory (correlation) time. If, on the other hand, a narrow spectral feature, relating to a long-lived vibrational mode, is excited then this will lead to long lasting quasi-coherent motion of the environment with the possibility of triggering quasi-coherent exchange with the system. This and other non-equilibrium exchanges between system and environment can have considerable influence on the system dynamics \cite{ChinHP12,KolliOS+12,ChinPR+13,BiggsC12,RichardsWCQ+12,WomickM09,WomickM11,WomickWS+12}, influence the direction of energy transfer \cite{ChinPR+13} and may even lead to the generation of long-lived quantum coherence in the system \cite{ChinPR+13,TiwariPJ13,AlmeidaHP+13,ChristenssonKP+12}.

The functional role of this non-equilibrium exchange and in particular the interaction of electronic system degrees of freedom with long-lived vibrational modes represents a key question in quantum biology. Thus the investigation of the non-equilibrium, non-perturbative and non-Markovian character of the system-environment interaction is of considerable interest. Besides the conceptual impact that it may have it also calls for the development of novel methods for the accurate numerical description of such system-environment interaction which must be capable of going well beyond the usual perturbative treatments in the description of most quantum technologies \cite{PriorCH+10,IshizakiF09a,MakarovM94,WeissET+08,HuoC10}.

{\em Control of Structure ---} The discussions so far suggest that quantum dynamics, structured environments and their mutual interplay may provide increased efficiencies and potentially even novel functionalities. In order to optimize this interplay between system and environment, biological systems cannot simply be random conglomerates but must possess structure that they can control. Examples are proteins which are essentially one-dimensional amino-acid sequences, representing the primary structure, that have the ability to fold into specific three-dimensional arrangements whose short range, secondary, structure defines the local arrangement via the relative orientation of the amino acid residues, while the global three-dimensional arrangement, denoted as the tertiary structure, provides the long-range structure. It will become crucial that far from being passive scaffolds whose mere purpose is to hold other molecules in their place, the primary, secondary and especially the tertiary structure of proteins play a much broader and active role for the emergence of quantum effects in biology. In fact, their purpose is to control and tune the properties of the actual molecular network, the structure of the environment in which this network is embedded and, crucially, the interaction and interplay between the two. The protein can achieve this by tuning properties such as local energies, by adjusting the local environment mainly via its secondary structure, and the coupling rates within the transport network, by adjusting the distance and relative orientation of constituents as determined mainly through its tertiary structure, and at the same time the properties of the environmental fluctuations that this transport networks are subjected to are determined and tuned mainly via the secondary protein structure. The latter may not only be achieved by structuring the protein itself but also by placing within it molecules that provide desirable vibrational features. This optimization that is required here will have taken place on evolutionary time scales. It is noteworthy though that, distinct from evolutionary optimization, via the control of conformation within an existing structure nature also possesses the means to switch between different states of operation.

How this control is realised in detail and which configurations exploit optimally quantum properties as well as system-environment interaction is of considerable interest for the exploration of quantum effects in biology.

{\em Technologies \& Experiments: Hardware and Software --} The questions and principles stated so far will guide our thinking and allow theory to progress. It must be stressed however that the study of quantum effects in biology is an endeavour in which theory alone cannot succeed and has to proceed hand-in-hand with experiment. So much is unknown about biological systems that assumptions that are entering theoretical models as well as the predictions of these models must be tested against experiment. Indeed, as we already stated in the introduction, the quest for quantum effects in biology has gained momentum recently in no small part due to the technological advances that are increasingly permitting the direct observation of quantum effects in biological systems and the probing of these systems on previously unknown length- and time-scales.

The future development of the field will depend critically on the identification and development of  technologies, classical or quantum physical, that can lead to deeper insights into the workings of biological systems at the length and time scales at which quantum effects can act. New technologies and new experimental set-ups as well as protocols need to be developed (see e.g. \cite{HarelFE10,CarusoSS+12,CarusoMC+12,HoyerCM+13} for a few examples). Furthermore, the complexity of the systems under investigation suggests that methods from fields such as signal processing \cite{CandesW08,AlmeidaPP12,SandersMS+12} may prove fruitful to optimize experiments and subsequent analysis.

It will be exciting to see how new experimental phenomena and theoretical principles will stimulate the development of novel experimental tools and clever protocols. This, we believe, will be essential for the development of the field as only irrefutable direct proof of quantum phenomena and their functional role will force biologists to adopt quantum theoretical concepts to understand foundational aspects of biology.

\section{Quantum Dynamics in Biological Environments}
Before we start to discuss a set of general and generalizable principles that govern the quantum dynamics of biological systems we would like to discuss here an example that presents us with a "smoking gun" which suggests that the study of the interplay between quantum dynamics in the presence of structured environments may lead to interesting insights.

\subsection{Transport \& Environment: An illustrative example}
Let us consider a transport network that is composed of a set of highly absorptive entities such bacteriochlorophyll molecules each of which may support an electronic excitation (an Frenkel exciton). These molecules will be arranged in space by a protein scaffold and together they form a pigment-protein complex. In our concrete example this will be the Fenna-Matthews-Olson (FMO) complex \cite{AdolphsR06,SchmidtamBuschMM+11} which forms an integral part of the photosynthetic light harvesting complexes of green sulphur bacteria \cite{vanAmerongenVvG00}. Such complexes serve to transport electronic excitations, excitons \cite{Davydov64}, in the presence of a vibrational environment.

We will contend that nature can construct and optimize pigment-protein complexes, for example the FMO complex, to create transport networks that exploit both the quantum dynamics of its electronic degrees of freedom and their interaction with a structured vibrational environment. The detailed mechanisms that nature has used to achieve this and the design principles it is following will be discussed in the next section and then used as the basis to understand other seemingly unrelated biological quantum phenomena.
\begin{figure}[t]
\centerline{\includegraphics[width=8cm]{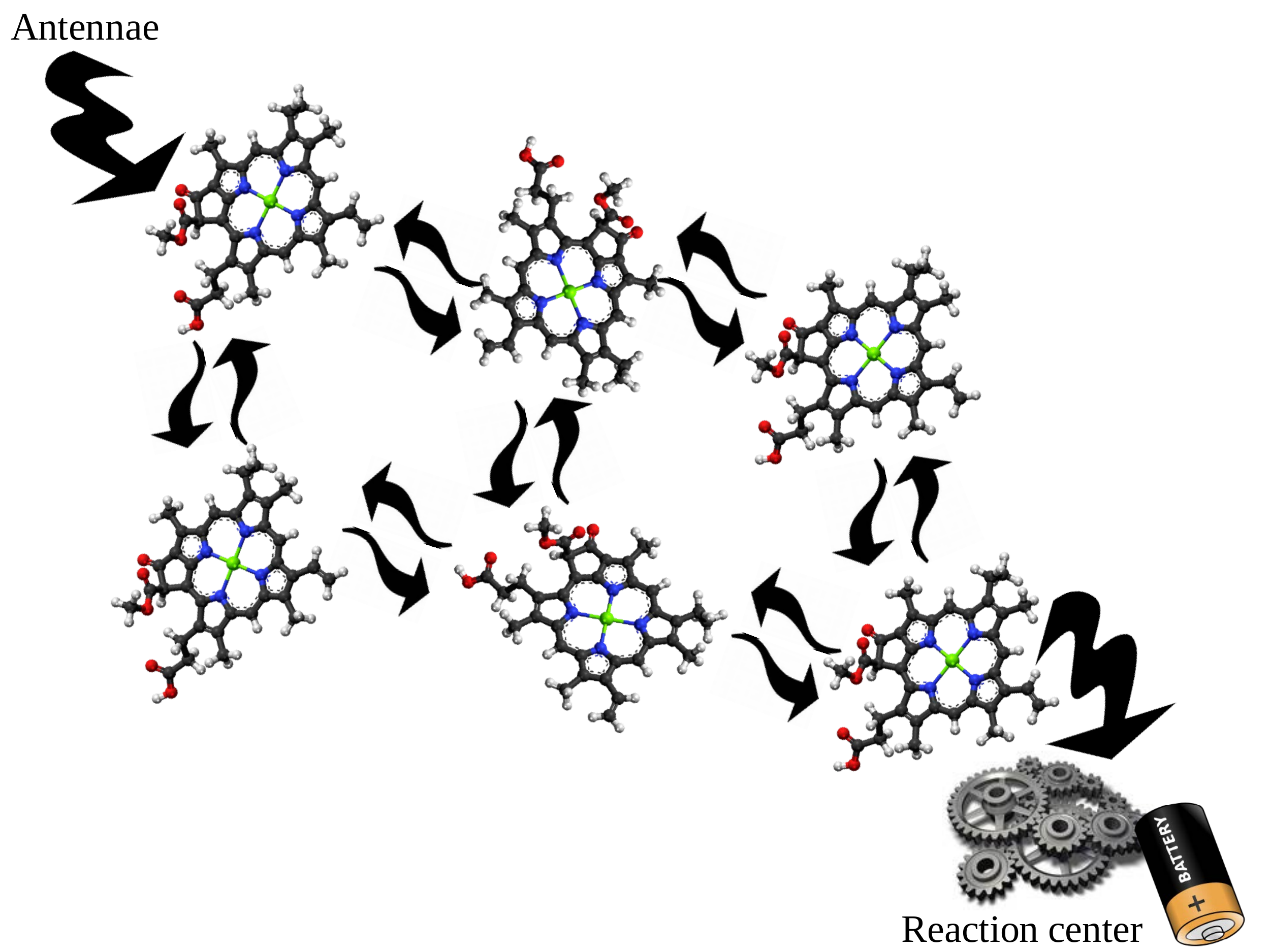}}
\caption{A schematic picture of the transport network such as the one realized in the Fenna-Matthews-Olson complex. Molecules, such as Bacteriochlorophyll a (BChla), are arranged in space giving rise to specific distances and relative orientations between individual BChla thereby adjusting the strength of the dipolar interaction (indicated by black arrows) between excitations on different BChla molecules. The surrounding protein (not shown) is also able to control the local environment of the BChla molecules and thus adjust their excitation energies. The specific arrangement and the nature of the environment determines the transport efficiency from the left upper site (site $1$ in the FMO) that accepts excitations from the antenna complex and transfers them via the lower right site (site $3$ in the FMO complex) to the reaction center where charge separation is initiated to begin the irreversibly process of binding the exciton energy in chemical form.}
\label{Network}
\end{figure}

Let us first provide evidence, by means of a numerical example, that such tuning and optimization may be taking place and that the vibrational environment may indeed have a significant and indeed beneficial impact on the transport dynamics of the FMO complex. Then we will move on to derive generalizable conclusion from these findings. The full Hamiltonian describing the exciton-vibrational interaction as well as the exciton-exciton interaction is given by $H = H_{ex} + H_{I} + H_{B}$ where
\begin{eqnarray}
    H_{ex} &=& \sum_{n=1}^{N} E_n |n\rangle\langle n| + \frac{1}{2}\sum_{m \neq n} [J_{mn}|m\rangle\langle n| + h.c.], \\
    H_B &=& \sum_{i,k} \hbar\omega_k a^{\dagger}_{ik} a_{ik}, \\
    H_I &=& \frac{1}{2} \sum_{n} [\sum_k \sqrt{S_{nk}} \omega_k (a_{nk} + a^{\dagger}_{nk})
    |n\rangle\langle n| + h.c.].
\end{eqnarray}
Here $|n\rangle$ describes an excitation on site $n$, the $J_{mn}$ describe the dipolar interaction between excitation on sites $m$ and $n$ and the operators $a_{nk}, a^{\dagger}_{nk}$ denote bosonic destruction and creation operators for the $k^{th}$ independent vibrational mode coupled to site $n$ \cite{AdolphsR06}. The exciton-mode interaction is determined by the strength of their Huang-Rhys factors $S_{nk}$ \cite{WendlingPP+00}. Note, that $H_I$ describes a purely dephasing interaction because the vibrational degrees of freedom have energies that are at least $10-100$ times smaller than the excitation energies of the excitons thus suppressing direct exciton-phonon interconversion. The dephasing can be understood to originate from the fact that vibrations will change the local environment of each site (e.g. moving charges) and thus affect the excitation energy of the relevant site \cite{MayK04}. We assume that the dynamics is dominated by contributions to the spectral density where each site interacts with its own independent environment an assumption that is corroborated by first-principles numerical studies of photosynthetic complexes \cite{StrumpferS11,OlbrichSS+11,OlbrichJL+11,OlbrichSS+11a,ShimRV+12} and normal-mode analysis combined with quantum chemical methods \cite{RengerKS+12,RengerM13}. The interaction between site $n$ and its environment is characterized by the spectral density $J_n(\omega) = \sum_k S_{nk} \omega_k^2 \delta(\omega - \omega_k)$ which is a joint property of the environment and the system combining the strength of interaction of modes with the mode density.

The full dynamical equation also need to include two further contributions, one describing the spontaneous annihilation of an exciton and the concomitant loss of the energy into the general environment at a rate $\gamma_{loss}$, a process that nature would like to avoid. Secondly, the transfer of the excitation from a specific site, in the FMO complex, that is the site labeled $3$, into the reaction center which is again described by an irreversible decay at rate $\gamma_{RC}$ motivated by the fact that in the reaction center charge separation is achieved to irreversibly stabilize the excitonic energy. Both contributions enter the equations of motion via typical Lindblad terms \cite{RivasH12} so that we find the global time evolution of the density operator describing system and environment to be governed by
\begin{eqnarray}
    \frac{d\rho_{S,E}}{dt} &=& -i[H , \rho_{SE}]\nonumber\\
    && \hspace*{-1.cm} - \gamma_{loss}\sum_{n=1}^N (|n\rangle\langle n|\rho_{SE} + \rho_{SE}|n\rangle\langle n| - 2 |0\rangle\langle n| \rho_{SE} |n\rangle\langle 0|) \nonumber \\
    && \hspace*{-1.cm} -\gamma_{RC}(|3\rangle\langle 3|\rho_{SE} + \rho_{SE}|3\rangle\langle 3|- 2 |0\rangle\langle 3| \rho_{SE} |3\rangle\langle 0|)
\end{eqnarray}
where $|0\rangle$ denotes the electronic ground state of the pigment-protein complex. Needless to say, these dynamical equations describing the full state of system and vibrational environment are generally too complex to be solved exactly. Crucially however we will see shortly that the structure of the environment does play an important role and thus the development of methods that can capture these features will be of considerable importance for quantitative studies of such systems. In order to bring out the main points clearly, we would like to delay the discussion of these numerical challenges to final part of this article. In order to obtain our first observation we follow the approximate treatments presented in \cite{MohseniRL+08,PlenioH08,CarusoCD+09,delReyCH+13} and will improve on these later on.

\begin{figure}[bt]
\vspace*{-3.cm}
\centerline{\includegraphics[width=8.5cm]{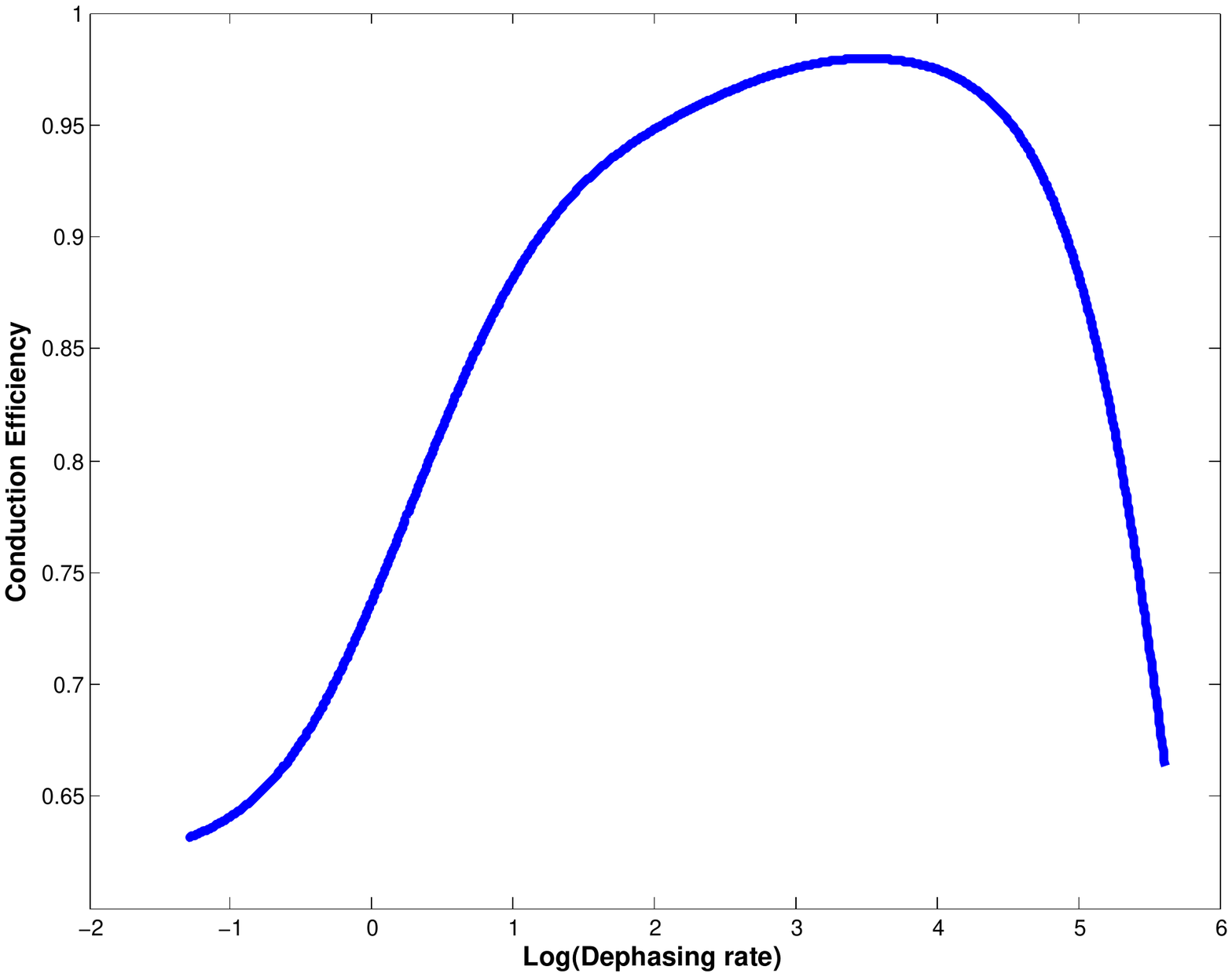}}
\vspace*{-3.cm}
\caption{Plot of the conductivity of the FMO complex where excitations enter the FMO complex
at site $1$ and exit at site $3$ versus the overall strength of the dephasing noise due to the interaction between electronic and vibrational degrees of freedom. The key observation is that optimal performance is found for an intermediate level of dephasing noise. No noise or very strong noise are counterproductive for the performance of the transport network. See \cite{CarusoCD+09} for the Hamiltonain parameters and the spontaneous emission rate, the noise is modeled as local dephasing Lindblad master equation in the site basis.}
\label{TransportvsNoise}
\end{figure}

To this end we will model the system-environment interaction perturbatively and derive a master equation for the system evolution only to model the transport through the FMO complex under low light conditions, i.e. the rate $\gamma_{in}$ at which excitations enter the network via site $1$ is much smaller than the rate $\gamma_{RC}$ at which excitation leave the network from site $3$ into the reaction center. As a consequence, the mean population in the transport network is much smaller than unity at any time. The typical but at first sight perhaps surprising result of such an analysis is presented in Fig. \ref{TransportvsNoise}.
Contrary to what one might have expected, the conductivity of the electronic transport network in the FMO complex (quantified as the rate at which the reaction center is populated in steady state divided by the rate at which excitations enter the transport network) exhibits a maximum at a finite dephasing rate, that is, dephasing noise can actually assist the electronic transport \cite{PlenioH08,MohseniRL+08}.

This immediately raises the question as to whether the regime that results in optimal transport performance is found to be essentially classical in the sense that the dynamics is well represented by a rate equation model or whether, despite the dephasing noise, it remains firmly in the quantum mechanical regime in which quantum coherent dynamics is only weakly perturbed by dephasing noise.
Questions of this type can be answered both in a qualitative and a more quantitative manner. First,
an examination of the parameters that are typically entering the dynamical equations when describing photosynthetic complexes reveals that the strength of the intra-system coupling, e.g. the dipolar interaction, is comparable in strength to the system-environment coupling. Hence one already expects that the dynamics is taking place in a regime in which neither dephasing noise nor quantum coherent dynamics clearly dominate. This is further corroborated by the examination of the coherence and entanglement \cite{PlenioV98,PlenioV07} properties of states \cite{CarusoCD+09,FassioliO10} and, more importantly, the dynamics of the system \cite{CarusoCD+10} which demonstrate quite clearly that on shorter length- and timescales quantum coherence is present in the systems while for longer distances and times classical properties dominate. This suggests that indeed, the optimal operating regime in this setting is found to be "halfway" between the classical and the quantum world.

In the light of recent discussions concerning the relevance of coherence in photosynthetic systems, we feel that an important remark is in order here. The coherence properties of the states and dynamics of a system that are {\em observable} in an experiment may depend very much on the specific experimental set-up and the specific excitation regime that the system is subjected to. Excitation by incoherent sunlight as well as ensemble averages may suppress the {\em observed} coherence and have tempted some researchers to reach the conclusion that coherence may be of no relevance in these systems under natural conditions. This however is not necessarily correct as the crucial point are the coherence properties of the underlying dynamics and not of signatures of coherence in an experimental signal. It is the equations of motion that determine the performance of each individual system and are unaffected by the nature of the initial preparation or ensemble averages that may obscure the observed coherence (consider a set of pendula each of which oscillates independently from the others at a fixed frequency and phase. If the phase for each pendulum is chosen at random then the global signal appears incoherent while clearly each pendulum is coherent).
While natural conditions do not resemble laser light, laser spectroscopic experiments on individual specimens provide the sharpest tools for the identification of the dynamical equations that govern the system evolution and thus have a crucial role to play in the determination of quantum effects in biological systems.

These observations raise the questions as to why optimal performance is achieved in this intermediate regime. Answering this question will lead us to identify the dynamical and structural principles that are underlying optimal performance of quantum transport networks. It will drive us towards uncovering a rich interplay between electronic degrees of freedom and their vibrational environment and point towards the possibility that nature has optimized both electronic networks and vibrational environment in an evolutionary process. This will be the subject of the next section.

\section{Design Principles}\label{Quantum}
Here we will elucidate basic principles that have been found to underlie the fruitful interplay between vibrational environments and coherent quantum dynamics
\cite{MohseniRL+08,PlenioH08,OlayaCastroLO+08,CarusoCD+09,CaoS09,CarusoCD+10,ChinDC+10,WuLS+10,HoyerSW10,RebentrostMK+09,MoixWH+11,PlenioH11,CamposVenutiZ11,delReyCH+13,ChinPR+13}.
Identifying and understanding these principles at a deep, intuitive level and seeing how nature may have used them to optimize performance does provide additional value even if individual processes in specific circumstances have been known in different physical situations. The principles that we present here will also allow us to bring under one umbrella several seemingly unrelated biological phenomena, namely excitation, electron and proton transfer processes, the chemical compass model of magneto-reception of birds and the mechanisms underlying olfaction, each of which is suspected to be governed in an essential way by quantum phenomena.

\subsection{Controlling resonances -- The phonon antenna}\label{phonon}
We will begin by elucidating a first principle which will provide an understanding why optimal transport performance in the FMO complex can be achieved at intermediate noise levels. More importantly, this principle is sufficiently general to provide a mechanism that can support the surprisingly long-lasting oscillatory features observed in recent ultra-fast laser spectroscopy experiments \cite{EngelCR+07,MercerEK+09,PanitchayangkoonHF+10,ColliniWW+10} and explain key aspects of the dynamics that may underlie the process olfaction.

\begin{figure}[hbt]
\centerline{\includegraphics[width=8cm]{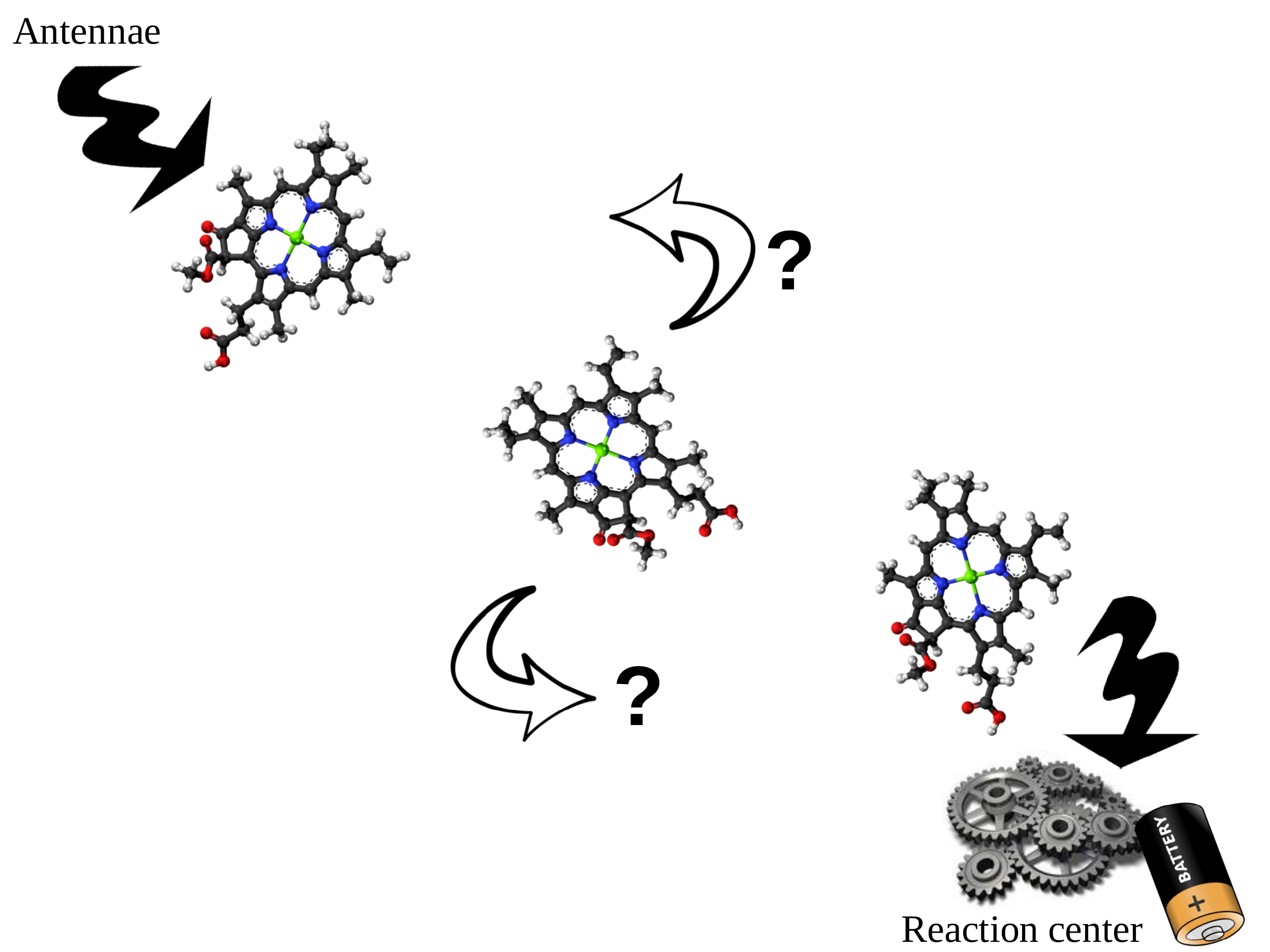}}
\caption{A network made up of only three sites, site $1$ which accepts excitation from the antenna, site $3$ which is connected to the reaction center both of which are fixed in their properties (position, orientation and excitation energy) as well as another site $2$ whose excitation energy, position and orientation, and hence dipolar interaction strength with sites $1$ and $2$ we are free to chose. What is the optimal choice of the excitation energy of site $2$ and its dipolar coupling strengths to sites $1$ and $3$?}
\label{Optimization}
\end{figure}

\begin{figure}[hbt]
\vspace*{-0.25cm}
\centerline{\hspace*{1.5cm}\includegraphics[width=11cm]{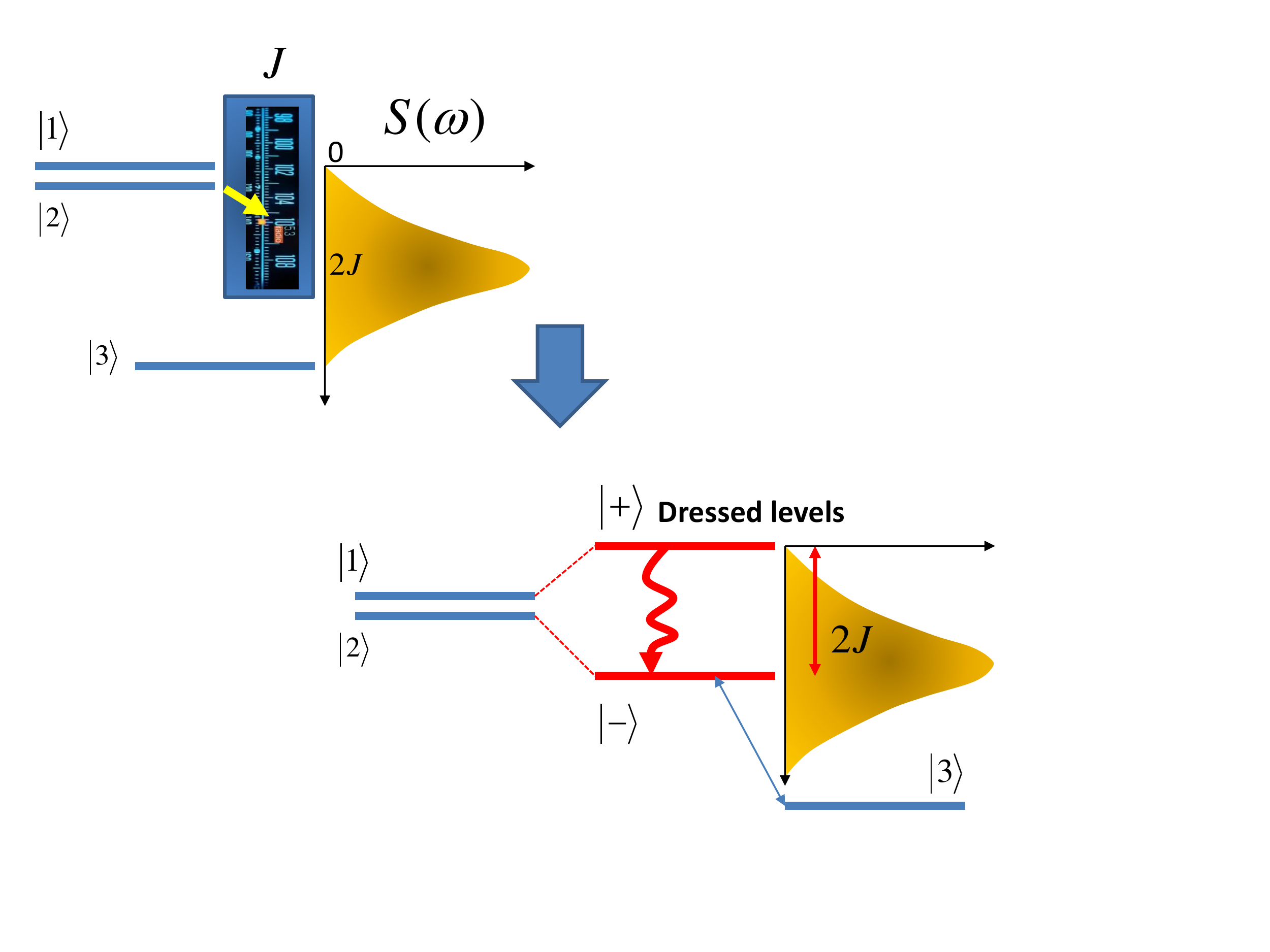}}
\vspace*{-1.cm}
\caption{In the upper figure, two closely spaced energy levels are separated from a third level to which excitations should be delivered. They are subject to dephasing noise from an environment with a finite bandwidth that exhibits a maximum. A coherent interaction between the upper two energy levels leads to dressed states $|\pm\rangle$ with an energy splitting which, if matched to the maximum of the environment spectral density, will optimize transport from the upper to the lower level. Hence the dressed states act as an antenna to harvest environmental fluctuations to enhance transport.}
\label{FigAntennae}
\end{figure}

We will approach this topic by means of a simple but instructive question concerning the optimization of a simple transport network (see fig. \ref{Optimization} for a schematic representation of the following). Consider a network made up of only three sites, namely site $1$, which accepts excitations from the antenna and site $3$ which is connected to the reaction center. Both, site $1$ and site $3$ are fixed in their properties (position, orientation and excitation energy). The system is completed by site $2$ whose excitation energy, position and orientation, and hence dipolar interaction strength with sites $1$ and $3$ we are free to choose. We assume that site $3$ provides the zero of excitation energy, site $1$ has an excitation energy that is $300$ wavenumbers higher (for readers unaccustomed to wavenumbers, note that $1$ wavenumber, also denoted as $1cm^{-1}$, corresponds to $\omega=1.88\cdot 10^{11} s^{-1}$). The question that we would like to answer concerns the optimal choice of excitation energy, position and orientation of site $2$ or in other words the optimal choice of the excitation energy of site $2$ and its dipolar coupling strengths to sites $1$ and $3$?
As such, this question cannot be answered unambiguously as we are missing a crucial piece of information, namely that of the structure of the environmental fluctuations. This structure is characterized by the spectral density of the environment which is a combination of the density of environmental modes and their individual coupling strength to the system. Typical spectral densities in pigment-protein complexes possess considerable structure with sharp peaks originating from long-lived vibrational modes as well as a broad background whose maximum tends to be in the range of around $200cm^{-1}$ which we will now assume for the subsequent optimization. For the sake of clarity, let us assume that the environmental spectral density has a single maximum and thus takes roughly the shape depicted in figure \ref{FigAntennae}. A numerical optimization employing Redfield equations to take account of the spectral structure of the environment (see \cite{ChinHP12,delReyCH+13} for theoretical and numerical details) now finds that the optimal position of site $2$ is close to site $1$ such that it exhibits a strong coherent dipolar interaction and close in excitation energy. Having found this numerical results we now would like to rationalize its origin and thereby arrive at a very useful design principle -- the phonon antenna.

Indeed, the strong coherent dipolar interaction between sites $1$ and $2$ suggests that we move to a new basis made up of the eigenstates of the coherent part of the dynamics of these two sites, that is the excitonic states of that system or, for quantum opticians, the dressed state picture. This change of picture leads us to rewrite the Hamiltonian eq. (3) that describes the system-environment interaction in the excitonic basis of eigenstates $\{|e_n\rangle\}$ of eq. (1), so that $|i\rangle=\sum_{n}C_{n}^{i}|e_{n}\rangle$, and the coupling terms
\begin{eqnarray}
    H_{I}  &=& \frac{1}{2}\sum_{n,m}(Q_{nm}|e_{n}\rangle\langle e_{m}|+ h.c.),\label{ht}
\end{eqnarray}
where
\begin{eqnarray}
    Q_{nm} &=& \sum_{ik}\sqrt{S_{k}}\omega_{k} C_{n}^{i}C_{m}^{i}(a_{ik}+a_{ik}^{\dagger})\label{r}.
\end{eqnarray}
This leads us to two insights. Firstly, in the exciton (dressed state) basis the action of the dephasing noise now leads to transitions between excitons, that is amplitude noise, which facilitates transport towards the lower of the two exciton states. Secondly, the two excitons (dressed states) are separated by an energy difference that is related to the coherent dipolar coupling strength and the energy difference of sites $1$ and $2$. The dominant contribution to the transition between these excitons (dressed states) arises from those environmental modes whose frequency closely matches the energy difference between dressed states. Indeed, the optimal solution is such that the energy separation of the dressed states matches the maximum of the environmental spectral density, i.e. where the environmental fluctuations are strongest, so that the environment may bring about transitions between the dressed states most effectively. In this sense, we can argue that the two eigenstates of the coupled Hamiltonian are tuned to harvest environmental fluctuations to achieve optimal excitation energy transport through the formation of a tunable "phonon antenna".

Observing these two points alone, that is inducing strong coherent coupling to move to a dressed basis and tuning the coupling such that it matches the maximum of the spectral density of the environmental fluctuations, one already comes close to the numerically obtained solution of the above optimization problem and can thus optimize excitation energy transport. It should be noted that the phonon antenna principle is also capable of making predictions about more complex transport networks such as that of the FMO complex. Indeed, it was found that the physically important relaxation pathway between sites $1$ and
$3$ is mediated by pigments which are spectrally and spatially positioned by the protein to efficiently sample the spectral function of the protein's fluctuations \cite{ChinHP12,delReyCH+13}. Whether this optimality is a determinant in the emergence of this structure in nature is another matter altogether, but it is striking how well the phonon antenna concept can be used to rationalise the site energies and couplings of the pigments participating in this pathway. The role of vibrational modes has also been studied for other light harvesting complexes such as the cryptophyte antenna protein phycoerythrin 545 \cite{KolliOS+12}.

It now becomes transparent why the optimal operating regime for excitation energy transport may actually be found to be where coherent dipolar interactions and system-environment interactions, i.e. dephasing noise, are of broadly comparable strength. Indeed, if the environmental noise is too weak, then the formation of dressed states will present little benefit for transport. On the other hand if dephasing noise is very strong then it will suppress the formation of the dressed states and thus of the phonon antenna effect in the first place. As a consequence, an intermediate regime where dephasing noise of intermediate strength is present naturally appears as the optimal operating regime according to the phonon antenna. We will later see that there are a number of additional mechanisms in which noise and coherent dynamics coexist to lead to similar conclusions \cite{CarusoCD+09,LloydMS+11}.

Before we move on to the discussion of these effects, it is instructive to highlight a connection that the phonon antenna concepts shares with the concept of Hartmann-Hahn resonances \cite{HartmannH62} in nuclear magnetic resonance and spin sensing. Here one is faced with the challenge that the sensor, e.g., an electron spin, may not be resonant with an external system generating a signal (e.g., a different electron or nuclear spin). This problem may be overcome by continuously driving the electron spin in the sensor at a strength that leads to dressed states whose splitting matches the frequency of the external signal and thus permits a strong coherent response of the sensor (transitions between upper and lower dressed states) \cite{HartmannH62,CaiRJ+13,CaiJP+13}.

With this in mind we now move on to bring out two other conclusions and connections that we can draw from the phonon antenna concept in the case when the spectral density of the environment possesses very sharply peaked features, that is well-defined long lived vibrational modes. Indeed, this will provide a both a mechanism explaining the origin of long-lived coherences observed in recent ultrafast spectroscopy experiments and a connection to biological sensors.

\subsubsection{Long-lived coherences as a non-equilibrium process} \label{Sec:longlived}
Experimental observations employing ultrafast 2-D spectroscopy on various photosynthetic complexes exhibited long-lived oscillatory features which were interpreted as evidence for long-lived electronic coherence in the systems under investigation \cite{EngelCR+07,PanitchayangkoonHF+10,ColliniWW+10}. Under this hypothesis electronic coherence appear to exhibit lifetimes that can reach the picosecond range thus exceeding expectations from condensed matter systems at least tenfold. This interesting observation gave rise to a variety of attempts for explanations of the long-lived coherences including (i) overall reduction of dephasing \cite{PachonB11,ShimRV+12} which are however not compatible with the observed very short lifetimes of optical exciton coherences in the system, (ii) correlations in the noise sources between different sites \cite{FassioliNO10,StrumpferS11,LimTY+13} which are however not supported by first principles calculations of spectral densities \cite{OlbrichSS+11,OlbrichJL+11,OlbrichSS+11a,ShimRV+12} and normal-mode analysis combined with quantum chemical methods \cite{RengerKS+12,RengerM13} and (iii) variations of the electronic structure of the FMO complex \cite{RitschelRS+11a} which are not sufficient however to explain the observed durations. In the following we will show that the inclusion of significant coupling of electronic motion to long-lived vibrational modes \cite{ChinHP12,ChinPR+13} are capable of explaining the observations \cite{TiwariPJ13,AlmeidaHP+13} and even more so to give support to the idea that vibrational motion plays and important role for electronic transport, quantum or classical -- a principle which we will show here to be of broader importance in biology.

To gain an insight into possible mechanisms that support long-lived electronic coherences in biological systems, we will now take the phonon antenna principle to its extreme by applying it to a system-environment interaction in which the broad features of the spectral density are supplemented by sharp features due to the presence of some long-lived and well-defined vibrational mode as proposed in \cite{ChinPR+13}. As we have learnt in the previous section, tuning the dipolar interaction in the electronic system to the energy difference of the excitonic states that matches the maximum of the spectral density will maximize the rate at which transitions between these states occur. For a broad and smooth spectral density these transitions will be dominantly incoherent (following essentially a Fermi's golden rule argument). In the presence of a long-lived vibrational mode the phonon antenna principle remains the same but, crucially, the nature of the interaction changes as the interaction between a single vibrational mode and the electronic degrees of freedom is coherent, at least for as long as the mode itself remains coherent.

\begin{figure}[hbt]
\centerline{\includegraphics[width=8cm]{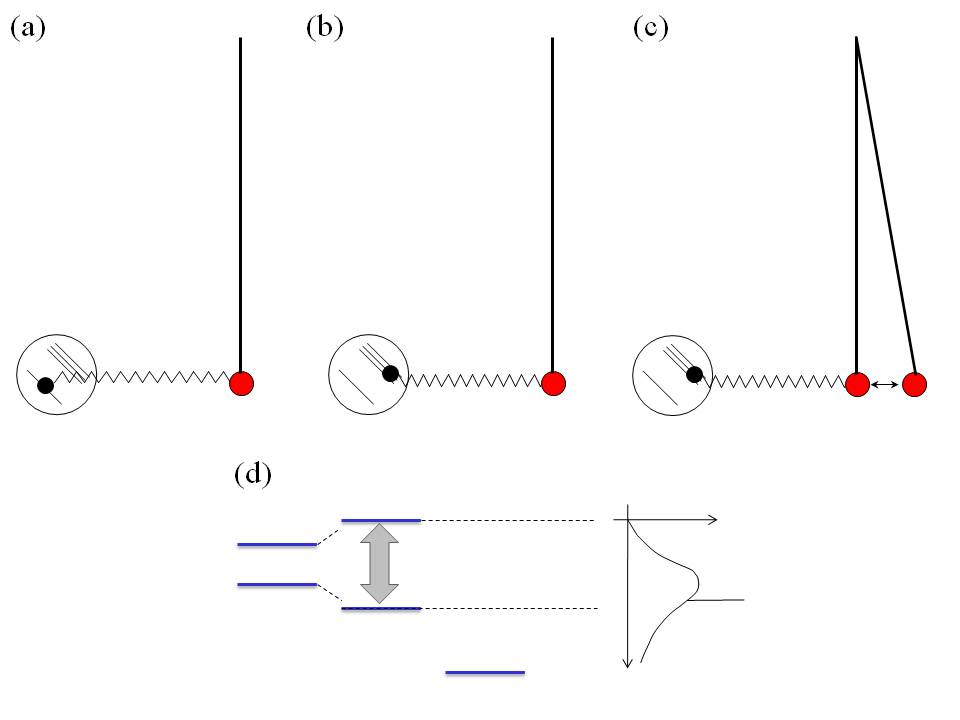}}
\caption{Simplified mechanical illustration of a possible principle behind long-lived coherence in biological systems. Details are presented in the main text.}
\label{longlived}
\end{figure}

Let us examine this mechanism in more detail by considering an exciton transport network in which excitons enter in one site and exit in another. Initially the network is in its ground state and no excitons are present. This situation is depicted schematically in Fig. \ref{longlived}a where a pendulum represents the long-lived vibrational mode which we assume to be in a thermal state with small excitation number and where the black dot represents the population of the ground state of the electronic system. The higher lying electronic levels, representing the various exciton eigenstates of the electronic system, are not excited even at room temperature as the excitation energy is in the range of eV.
The initial (fast) injection of an exciton, {\em either} coherently or incoherently, populates one of the exciton states of the system (raised black dot in Fig. \ref{longlived}b) and creates a sudden force on the
electrons and nuclei and thus change their equilibrium positions (compressed spring in Fig. \ref{longlived}b). Now the environment will start to react to these forces which initiates transient oscillations of the modes at approximately their natural frequency $\omega_k$. The continuous background of the spectral density will relax very rapidly into the new equilibrium state as it contains a broad range of frequencies and thus possesses a very short correlation time. The well-defined long-lived vibrational mode  will oscillate for a considerable time (which can be up to several picoseconds) and will interact with the electronic system (in Fig. \ref{longlived}c we see that the spring connecting the vibrational mode to the electronic motion is periodically expanded and compressed). This in turn leads to oscillations between different exciton states. We make two observations: Firstly, these oscillations will have the largest amplitude between those exciton states whose energy difference is nearly resonant with the frequency of the vibrational mode (see Fig. \ref{longlived}d).
Secondly, we note that the sudden displacement of this mode implies that it is now found to be in a displaced thermal state which will be close to a coherent state if the frequency of the mode is such that its thermal occupation number is low. As is well-known in quantum optics a mode in a coherent state acts on a two-level system essentially like a time-dependent classical driving field resulting in coherent Rabi-like oscillations following approximately the Hamiltonian
\begin{equation}
    H_{driving}\approx\frac{1}{2} \sum_{n\neq m}(\langle Q_{nm}\rangle (t)|e_{n}\rangle\langle e_{m}|+
    h.c.),
\end{equation}
where $\langle Q_{nm}\rangle (t)\propto \sum_{ik}\sqrt{S_{k}}\omega_{k} C_{n}^{i}C_{m}^{i}\sin(\omega_k t)$ in the above-mentioned approximation of the initial, transient and coherent response of the modes to exciton injection. As a result we will observe coherent transitions between dissipative excitonic states and thus coherences that will last for the coherence time of the vibrational mode -- an expectation that can be corroborated by more sophisticated numerical treatments employing methods described in \cite{PriorCH+10,ChinRH+10,ChinHP11,ChinPR+13}. This electron-vibrational coupling regenerates electronic coherences in the system to replace those that are continuously damped out by the fluctuations of the smooth background environment and may even lead to stationary entanglement \cite{HuelgaRP12}. We note that in the language of chemical physics, this phenomenon arises through coherent non-adiabatic coupling (via discrete mode motion) which induces oscillatory crossing of potential surfaces. The physical picture presented here illustrates a key point: electronic coherence may emerge from transiently exciting robust, weakly dephasing vibrational coherences which then transfer back coherence to those exciton transitions that are well matched to the mode \cite{ChinPR+13,TiwariPJ13,AlmeidaHP+13}.

The importance of vibrational modes for interpreting experimental observations in multidimensional spectroscopy has only recently begun to be appreciated \cite{PriorCH+10,ChinPR+13,ChristenssonKP+12,TiwariPJ13,AlmeidaHP+13} and from the discussion above is seen to be intimately related to the phonon antenna concept and in fact quantum sensing.

\subsubsection{Phonon-assisted electron tunneling \& olfaction}
In the following we will explain how the coupling of electronic and vibrational motion may also underlie the function of biological sensors and exemplify these ideas at the hand of a mechanism suggested to be an important contribution to the function of olfactory sensors.

Despite considerable progress concerning the understanding of the structure of olfactory receptors that involved in the early stages of the olfactory process, the detailed mechanisms by which we are able to discriminate between the vast number of odorants are not yet fully understood \cite{Zarzo07}. This is emphasized by the fact that for nearly $100$ years researchers have striven, with limited success, to identify principles that allow for the prediction of smell. The principal reason for this failure can be traced back to the lack of a detailed understanding of what is actually happening during and shortly after the process of binding of odorants to the bindings site of receptors. There are currently two mechanisms proposed that are based on quite different mechanisms and which, crucially, lead to different experimental predictions. It is possible that both contribute in a critical way to olfaction.

On the one hand there is the idea of the lock-and-key principle. The olfactory stimulus is provided my molecules, odorants, that arrive at the receptor via diffusion through the air. In the absence of the odorant the binding pocket and the receptor exhibits small thermal fluctuations about some equilibrium conformation. Only certain types of odorants will be capable of attaching themselves to the binding pocket, a choice determined by chemical affinity, shape etc (this is the "key in the lock" part of the principle). Once attached, the interaction between the odorant and the receptor results in a change of the average conformation of the receptor. This conformational change is then posited to induce further processes and initiates a signalling chain (this part constitutes the "turning of the key" part of the principle). For many receptors, especially those binding only a very specific molecule, this appears to be a useful and valid principle. Therefore it seems natural to adopt the very same principle also for olfactory receptors. There are at least $100000$ odorants but far fewer olfactory receptors -- several hundreds in humans, so that there is not a specific receptor for each single odorant. The ability to differentiate such a large number of odorants would thus require that each odorants may bind to a variety of receptors. This would give rise to a vast number of distinct binding patterns and the subsequent sensation of smell.

Despite its attractiveness and applicability in certain cases the lock-and-key principle has to be subjected to experimental test. This is where it has recently experienced some significant challenges. Firstly, it is not straightforward to explain by means of the lock-and-key mechanism alone why outwardly very similar molecules may smell completely different while molecules with rather different shape may smell similar. Even more remarkable is the observation in recent experiments that Drosophila flies are capable of discriminating between molecules in their hydrogenated and their deuterated form and, importantly, are able to generalize from deuterated molecules to other molecules that exhibit a vibrational modes similar in frequency to the Carbon-Deuterium stretch mode \cite{FrancoTM+11} (see \cite{GaneGM+13} for recent experiments with similar outcomes on humans).

Indeed, these observations provide some support for an alternative theory that is based on physical properties of molecules rather than chemical or shape-based ideas that are underlying the lock-and-key principle. Remarkably, it was well before the advent of the lock-and-key principle that Dyson in 1938 proposed \cite{Dyson38} that the smell of a molecule may be determined by its vibrational spectrum and that hence the olfactory system effectively operates as a vibrational spectrometer (and idea that has also been pursued later, starting in the 1950's, by Wright \cite{Wright77}). This is an attractive idea, especially to a physicist, but at the time it suffered from a quite severe drawback -- the lack of a plausible mechanism by which the olfactory system would in fact be able to identify specific vibrational modes. It was Luca Turin \cite{Turin96} who promoted the idea that inelastic electron tunneling (IET) may play a role in olfaction which, if true, would have the attractive feature that it would help to predict the smell of a molecule from the analysis of its vibrational spectrum \cite{Turin02}. IET is in fact a well established physical method for determining the vibrational spectrum of molecules \cite{HihathT12}. It was discovered in solid state physics by Jaklevik and Lambe in 1966 when they identified anomalous behaviour in the conductivity through tunnel junctions in the presence of organic adsorbates \cite{JaklevicL66,LambeJ68}. They realised that these anomalies arose for certain voltages such that the energy difference of the electrons on both sides of the tunneling barrier would match a vibrational quantum of energy of the adsorbed molecules and went on to demonstrate that this mechanism, inelastic electron tunneling spectroscopy, allows them to identify vibrational spectra of molecules (apparently, they recognized its potential for the theory of smell but were dissuaded by less far-sighted colleagues \cite{TurinPrivComm}).

\begin{figure}[hbt]
\centerline{\includegraphics[width=10cm]{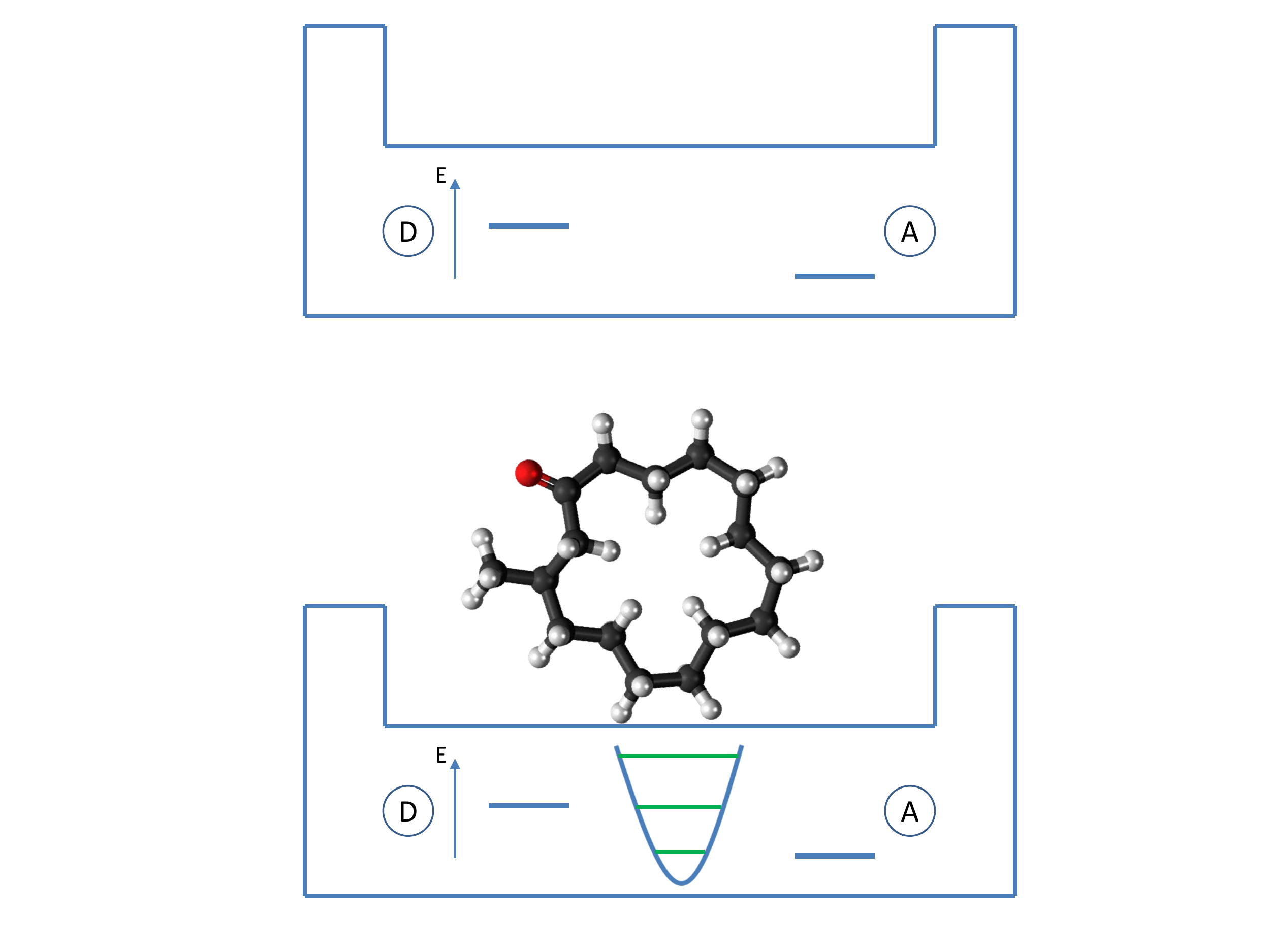}}
\caption{A schematic representation of the principle behind phonon assisted electron tunneling as a basis for olfaction. A binding site is close to a electron donor and acceptor that are at different energies which suppresses tunneling. If a molecule with a vibrational mode whose energy closely matches the energy difference between donor and acceptor enters the binding site then electron tunneling can become energetically allowed. Then a charge redistribution becomes possible and a subsequent signalling chain may be triggered.}
\label{olfaction}
\end{figure}

The proposed phonon assisted electron tunneling mechanism for olfaction makes use of two phenomena which, together, make it a proper quantum biological phenomenon. There is on the one hand the tunneling process of a massive particle, here the electron, and on the other hand the fact that a vibrational mode, that is a quantized harmonic oscillator, can only take up energy in discrete quanta proportional to the relevant vibrational frequency $\omega_{odor}$. It is the second aspect that makes it possible for this process to discriminate effectively between different vibrational modes and thus between the vibrational fingerprints of different molecules. A schematic picture of the process is presented in Fig. \ref{olfaction}. The binding pocket of an olfactory receptor, which begins empty (upper part of Fig. \ref{olfaction}), is assumed to be situated close to an electron donor and an electron acceptor (these may either be oriented alongside the binding site or separated by the binding pocket itself, a difference that is important in practice but not relevant for the following basic argument). Crucially, it is assumed that electrons in the donor have an energy that is $\Delta E$ higher than for electrons in the acceptor and that the energy distribution in donor and acceptor is much narrower than $\Delta E$. In that case tunneling is suppressed in the absence of the odorant because of the impossibility to satisfy energy conservation.
If the energy difference $\Delta E$ is well matched to the excitation energy of a specific vibrational mode $\hbar\omega_{odor}$ of the odorant, then its presence may facilitate a detectable and very specific change in the dynamics of the receptor by the following process: The odorant enters the binding pocket
and structural changes may take place which give space for the lock-and-key principle to play a certain role. Once the odorant has entered the binding pocket it will reside there for a certain time long enough for electron transfer to take place. The presence of the odorant with a vibrational mode that is well-matched to $\Delta E$ (see lower part of Fig. \ref{olfaction}) then allows for energy conservation during a tunneling process -- an electron can now tunnel from donor to acceptor while using its excess energy to excite a vibrational quantum of the odorant. This process will be unidirectional for sufficiently low temperatures such that $kT\lesssim \hbar\omega_{odor}$. The concomitant charge redistribution then triggers a further sequence of events that leads to a signalling cascade.

Needless to say very detailed calculations are required to ascertain that the described mechanism is possible in principle for reasonable choices of parameters \cite{BrookesHH+07,SolovyovCS12}. A wide variety of considerations need to be taken account in these studies which include the determination of spectral densities describing the electron-vibration coupling, the effect that orientation of the odorant in the binding pocket may have on it \cite{BittnerMC+12}, temperature effects etc (see \cite{BrookesHS12} for a more detailed analysis and \cite{Brookes11} for an overview).

What we would like to stress here though is the very close resemblance of the above mechanism to the phonon antenna principle and thus the importance of electron-vibrational coupling. In both cases an electronic degree of freedom senses the presence of a vibrational mode due to the tuning into resonance of the energy difference of two states (dressed states in the phonon antenna and donor/acceptor in olfaction) to the energy of single quanta of the vibrational motion. This resonance condition ensures increased transport efficiency and thus leads to a detectable effect at the physiological level (delivery of an electronic excitation to the reaction center or electron transfer which stimulates a subsequent response of a receptor). Both examples demonstrate the potential usefulness of controlled resonances for biological systems.

\subsection{Noise assisted processes}
Needless to say, not all biological systems will possess long-lived vibrational modes that are strongly coupled to electronic degrees of freedom and can therefore play a role in the dynamics of the system. What is present in all biological systems however are thermal fluctuations of the molecular and protein structures as well as the surrounding solvents, water etc. These may lead to a broad noisy background mainly resulting in dephasing noise on the electronic degrees of freedom.

Given the omnipresence of these noise sources one may ask as to whether nature has evolved to make use of in conjunction with quantum coherent dynamics to support processes that are of relevance to life. That this may indeed be the case we will explain in the following by first presenting several mechanisms by which dephasing noise may in fact support transport phenomena. This will be followed by a discussion of two phenomena in which noise can be shown to assist fundamental biological processes. In the first example we will briefly revisit transport in photosynthetic complexes while in the other example we will show that the mechanism that is proposed to underlie the magneto-reception of birds depends crucially on the presence of an environment.

\subsubsection{Bridging energy gaps \& blocking paths}
\begin{figure}[hbt]
\vspace*{-0.25cm}
\centerline{\includegraphics[width=9cm]{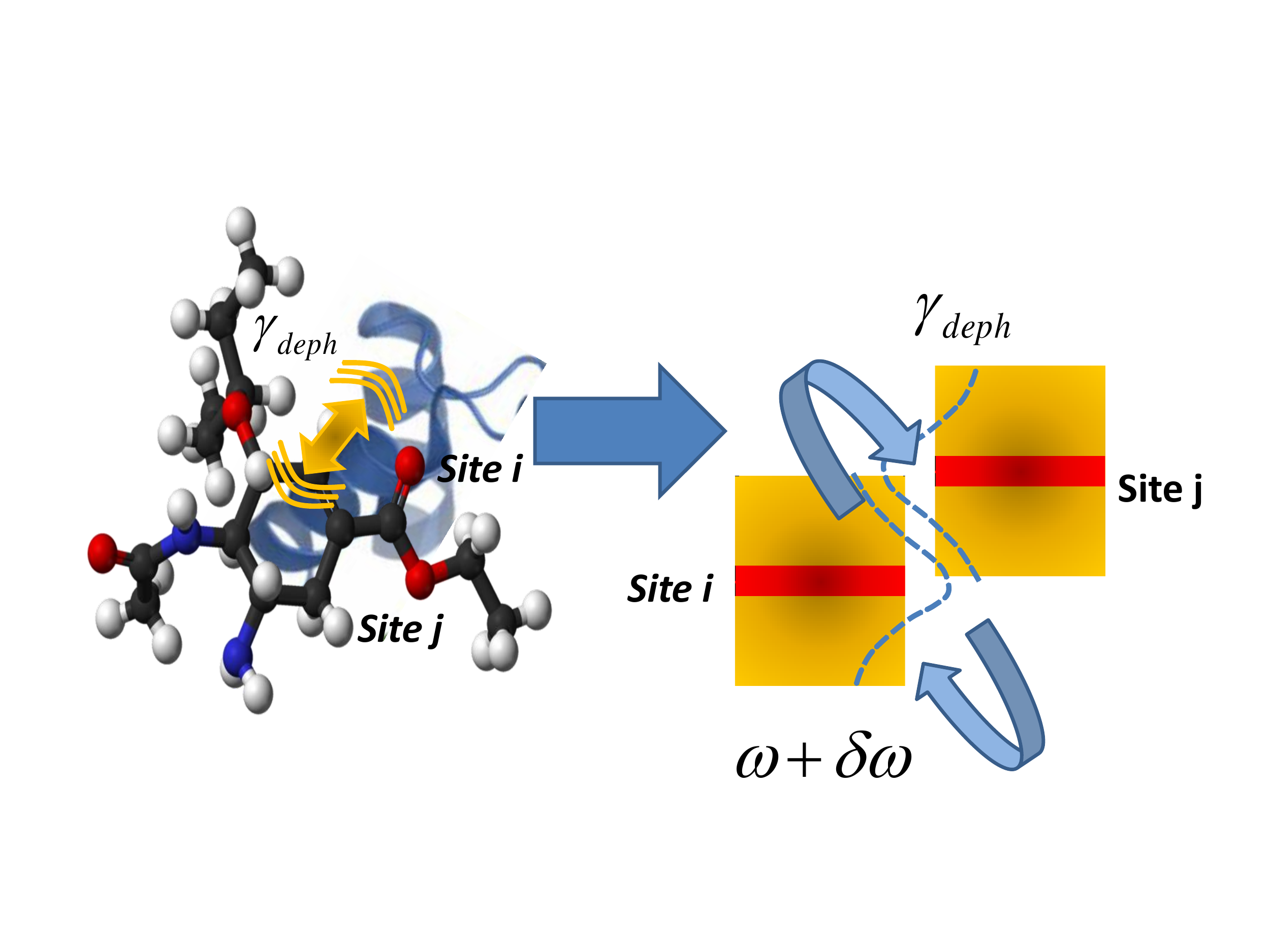}}
\vspace*{-1.cm}
\caption{Local dephasing, for example due to random fluctuations of the energy levels generated from random vibrational motion, leads to line-broadening and hence increased overlap between sites. Viewing these fluctuations dynamically, as illustrated by the double arrows, one finds that the energy gap between levels varies in time. The resulting nonlinear dependence of the transfer rate on the energy gap may therefore lead to an enhancement of the average transfer rate in the presence of dephasing
noise.
}
\label{Fig1}
\end{figure}

Pigment-protein complexes consist of a number of sites, schematically depicted in Fig. \ref{Network}, whose excitation energies generally exhibit a certain degree of static disorder, that is, their on-site energies will differ from site to site and also from one pigment-protein complexes to another. If the energy difference between sites that exchange excitation is larger than the intersite coupling matrix element in the relevant Hamiltonian, then transitions will be severely reduced because of energy conservation unless of course there is are quasi-resonant vibrational modes present that, as was explained in the phonon-antenna mechanism, can take up the energy difference. The presence of a well-matched mode is not necessary though.
Broadband dephasing noise alone may already come to the rescue in a manner that can be understood from two different viewpoints. On the one hand one notices that dephasing noise will lead to a broadening of the excitation energy of each site and thus to an increased overlap between the two energy levels while it does not cause the loss of excitations from the system (see Fig. \ref{Fig1}). Alternatively, one may take a dynamical viewpoint of the same phenomenon by realising that dephasing noise arises from the random fluctuations of the excitation energies of each site. As a consequence, these fluctuating energy levels will occasionally come sufficiently close in energy to allow for excitation energy transfer between the sites as the energy difference has been reduced to a value smaller than the direct coupling matrix element (see Fig. \ref{Fig1}). Again, we observe that this mechanism will lead to an optimal operating regime at intermediate levels of environmental noise. Indeed, a low level of fluctuations will not bring the site energies sufficiently close and transport remains suppressed, while excessive fluctuations of the site energies will reduce the time intervals in which the sites are energetically sufficiently close to allow for efficient energy transfer. This can be estimated easily by computing the overlap between
Lorentzian lines of width $\gamma$ that are displaced by an amount $\omega_0$ in which case we find
\begin{equation}
    \frac{1}{\pi^2}\int_{-\infty}^{\infty} \frac{\gamma}{\gamma^2 + \omega^2}
    \frac{\gamma}{\gamma^2 + (\omega-\omega_0)^2} d\omega = \frac{2\gamma}{\pi(4\gamma^2 + \omega_0^2)}.
\end{equation}
which takes on a maximum at $\gamma=\omega/2$.

An analogous behaviour arises already in the Marcus theory of electron transport where the rate constant $k$ for electron transfer reaction is given by
\begin{equation}
        k = A e^{-\frac{-\Delta G}{k_B T}},
\end{equation}
with $\Delta G$ given by
\begin{equation}
        \Delta G = \frac{\lambda}{4}\left(1 + \frac{\Delta G^0}{\lambda}\right)^2
\end{equation}
which achieves a maximum for $\Delta G^0=\lambda$. Here $A$ is a system dependent term, $\Delta G^0$ is the standard free energy of the electron transfer reaction and $\lambda$ is the reorganization energy which quantifies the strength of interaction between the electron and its environment composed of vibrational modes and the solvent \cite{Marcus93}.

It is worth noting that while the application of an excessive amount of dephasing noise suppresses transport this may in itself serve a useful function if for example a transport network contains sites, for example for structural reasons, that may lead to leakage of excitations into domains from where further transport may be slow. In such a case dephasing may be useful to reduce the effective transition rate to such sites and thus block unfavorable transfer paths from being followed \cite{ChinDC+10,ChenLL+13}.

\subsubsection{Destructive interference, symmetry and noise}
While linear networks can already exhibit interesting noise assisted transport phenomena \cite{GaabB04,PlenioH08,SemiaoFM10,AsadianTG+10,VaziriP10,KassalA12}, multisite networks may exhibit more complex behaviour which arises due to the interplay between a wealth of constructive and destructive interference effects in a quantum dynamical system on the one hand and environmental noise on the other hand \cite{PlenioH08,OlayaCastroLO+08,CarusoCD+09,CaoS09,CarusoHP10,GiordaGZ+11,MohseniSL+12,WuSC13}.

\begin{figure}[hbt]
\vspace*{-1cm}
\centerline{\includegraphics[width=10cm]{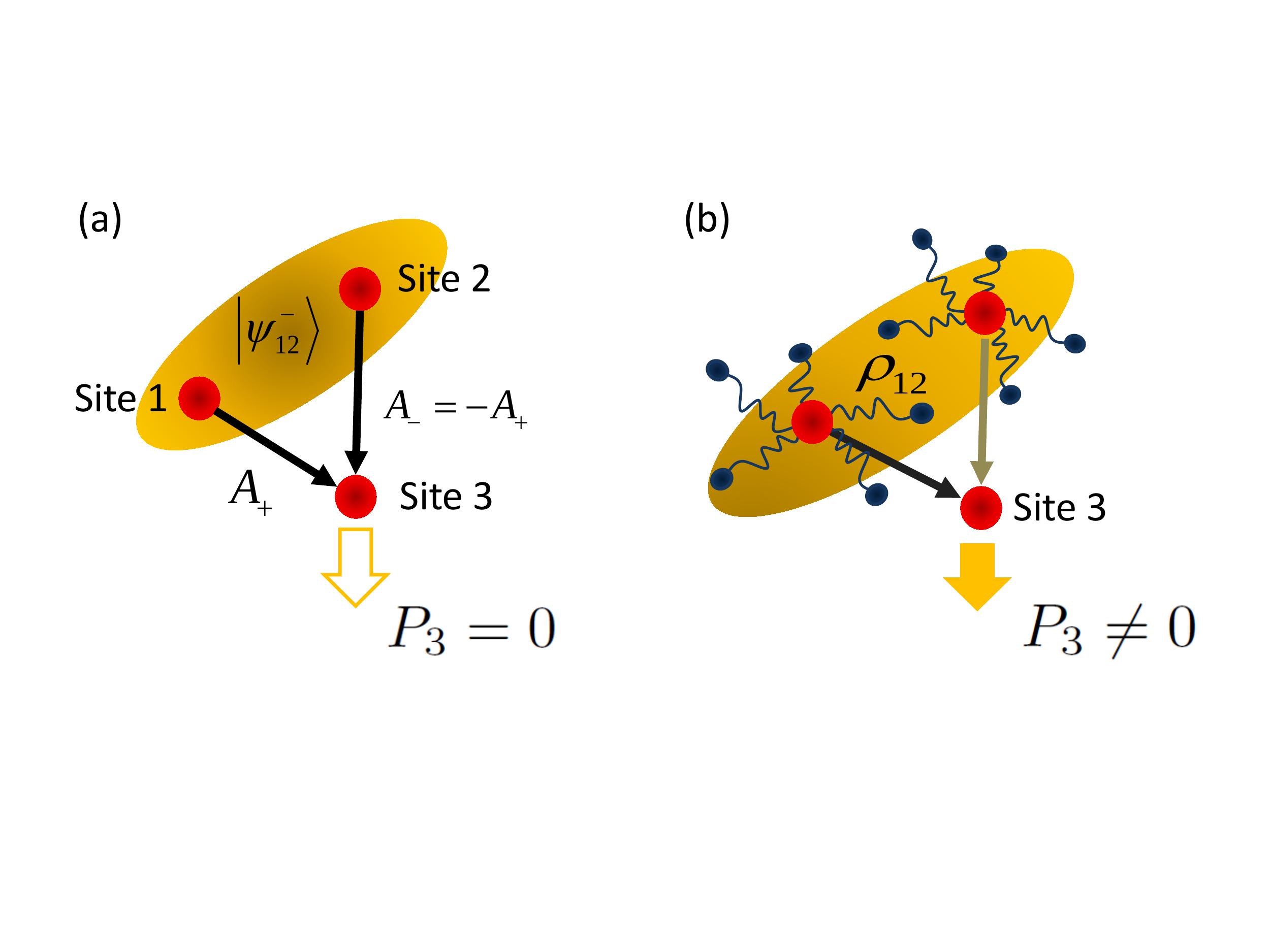}}
\vspace*{-2.cm}
\caption{
A three-site network in which two sites $1$ and $2$ are each coupled to a third site $3$ via an
exchange interaction of the same strength. Site three is irreversibly connected to a sink. In (a)
the excitation is delocalized over two sites (red and green) with equal probability of being found
at either site but with a wave function that is antisymmetric with respect to the interchange of red
and green. This state will not evolve due to destructive interference and hence no excitation will
ever reach the reaction center. In (b) pure dephasing causes the loss of phase coherence and the two
tunneling amplitudes no longer cancel, eventually leading to a complete excitation transfer to the sink.}
\label{Fig1a}
\end{figure}

A basic example that exhibits the essential nature of this type of effect consists of a simple three-site network depicted in Fig.\ref{Fig1a}. Here two sites $1$ and $2$ are coupled to a third site which in turn leaks excitations irreversibly into a reaction center. The coherent interaction is described by a Hamiltonian
\begin{equation}
    H = \sum_{k=1}^{3} E_i|i\rangle\langle i| + \sum_{k=1}^2 J_{k3} (|k\rangle\langle 3| + h.c),
\end{equation}
where $|i\rangle$ corresponds to an excitation in site $i$ and we assume $J_{13} = J_{23}$. Let us begin by considering an excitation initially prepared in the antisymmetric state
\begin{equation}
    |\psi\rangle = \frac{1}{\sqrt{2}}(|1\rangle - |2\rangle)
\end{equation}
which forms an eigenstate of this Hamiltonian whose overlap with the site $3$, which we assume to be coupled dissipatively to a reaction center, vanishes. As a consequence this excitation will remain localized and does not propagate through the system. Eventually the finite lifetime of the excitation implies that it will be lost to the general environment due to a spontaneous annihilation process, an event that is not in the interest of a transport network. Hence coherence effects may lead to a strongly reduced or even vanishing transport rate.

One may argue however, that under natural conditions a pigment-protein complex is not excited in such an antisymmetric state, but will tend to receive a single excitation locally, for example on site $1$ (this is for example the case of the FMO complex \cite{AdolphsR06}). Nevertheless it is easy to see that the subsequent dynamics has a propensity for leaving the system in an anti-symmetric state or more generally a state that propagates slowly through the network due to quantum interference \cite{CarusoCD+09}. To this end note that we can write the initial state localised on site $1$ as an equally weighted coherent superposition of the symmetric and the anti-symmetric states, i.e.
\begin{equation}
    |1\rangle = \frac{1}{\sqrt{2}}\left(\frac{|1\rangle - |2\rangle}{\sqrt{2}} +
    \frac{|1\rangle + |2\rangle)}{\sqrt{2}}\right).
\end{equation}
Thanks to constructive interference a symmetric superposition experiences a coherently enhanced coupling to the site $3$ to which it will then propagate rapidly, and from there into the reaction center, while the antisymmetric part will not evolve at all. Hence, in $50\%$ of the cases the system will remain in the anti-symmetric state while in the other $50\%$ of the cases the excitation reaches the reaction center. Therefore the transfer efficiency is limited to $50\%$ in this setting.

Now it becomes evident that dephasing noise whose strength is correctly tuned can have a beneficial effect in such situations. Indeed, uncorrelated dephasing noise acting locally on each site will randomly flip the relative phase between $|1\rangle$ and $|2\rangle$ and thus leads to transitions between the symmetric and the antisymmetric state. Hence, the presence of dephasing noise inhibits both constructive and destructive interference and therefore slows the propagation of an excitation in a system initiated in the symmetric state and accelerates the propagation of an excitation in a system initiated in the anti-symmetric state. As a consequence we expect again that an intermediate noise level will be optimal, too low it will not suppress destructive interference efficiently and too high it will suppress all transport. Noise will be beneficial if the overall propagation under the noisy environment is still sufficiently rapid to be completed within the natural lifetime of the excitation. Similar considerations have also been conjectured independently to play a role in biological electron transport in photosynthetic reaction centers \cite{BalabinO00} (note however that recent experimental and theoretical work suggests that long-lived nuclear vibrations are present in the reaction center dynamics and may have a role to play here too \cite{VosRL+93,VosJH+94,NovoderezhkinYv+04,StreltsovYS+96,WestenhoffPE+12,RomeroAN+13,RozziFS+13}).

This is the simplest example for the more general phenomenon that complex quantum networks may possess subspaces whose members do not propagate and do not have overlap with the site connected to the reaction center. Systems in which such a subspace will be populated during the time evolution, e.g. because it includes the site that receives energy, will benefit from environmental noise which can drive the systems out of the trapping subspace \cite{CarusoCD+09}.
It should also be noted that an energy mismatch and the resulting time-evolution of the relative phase between the two sites $1$ and $2$ also leads to transitions between symmetric and anti-symmetric state and can thus assists transport \cite{PlenioH08,CarusoCD+09,LimRL+11}. Which of the two processes if any plays a role in the actual dynamics of the the FMO complex needs to be determined by carefully designed experiments. In both cases the key point is that some process breaks the symmetry of the system and thus inhibits the system from getting trapped in an unfavorable state. It is this principle of breaking symmetries that will also play a key role in another sensory process in nature, namely the magnetic sense of birds as we consider shortly.

\subsubsection{Robustness of excitation energy transport in noisy quantum networks} As we have seen earlier in Fig. \ref{TransportvsNoise}, noise does indeed support transport through realistic transport networks such as that of the FMO complex. We have identified a variety of mechanisms for white as well as coloured noise to lead to these observations. Here we wish to highlight another aspect of noise assisted transport originating from the broad part of the spectrum, namely that it confers a certain stability on transport dynamics. In this case, neither variations of the fine details of the spectral density nor variations of the electronic structure of the transport network have a significant effect. As shown in \cite{CarusoCD+09,WuLM+12} transport networks that exhibit efficient white noise assisted transport will suffer relatively small variations of the transport efficiency, in the sub-percent range, even if the electronic parameters of onsite and coupling energies vary by up to $20\%$. Such a stability can be an attractive feature to ensure robustness in an organism that is subject to changes in its environment.

\subsubsection{Magneto-reception in birds}
The effects of weak magnetic field on the growth of plants as well as the remarkable orientation and navigation abilities of birds, mammals, reptiles, amphibians, fish, crustaceans and insects are well documented \cite{JohnsonL05}. The mechanism by which this magnetic field sense is achieved, however, is less well understood and at least two alternative ideas are being considered, one that exploits the forces exerted on ferrimagnetic iron oxide particles embedded in the body, essentially a classical effect, and another that is based on magnetically sensitive free radical reactions \cite{SchultenSW78,RitzAS00,Ritz11}.

Behavioral experiments with birds such as the European robin to study avian magneto-reception \cite{WiltschkoWR11} have led to the observation that the process of avian magneto-reception depends on the wavelength of the ambient light \cite{PhillipsB92} and can be disrupted by very weak external oscillating magnetic fields \cite{RitzTP+04,ThalauRS+05}. This together with the experimental demonstration of magnetic field effects on a radical pair reaction at typical earth magnetic fields \cite{MaedaHC+08} provide pieces of evidence to support the idea that the chemical compass mechanism may be involved in avian magneto-reception. The cryptochromes in the retina of migratory birds provide a potential physiological implementation of such a mechanism \cite{RitzAS00} which has motivated the recently growth in attention from quantum physicists \cite{Ritz11,Kominis09,CaiGB+10,GaugerRM+11,Cai11,CaiCP11,BandyopadhyayPK12,GaugerB13,HogbenBH12,CaiP13}.

\begin{figure}[hbt]
\centerline{\includegraphics[width=9cm]{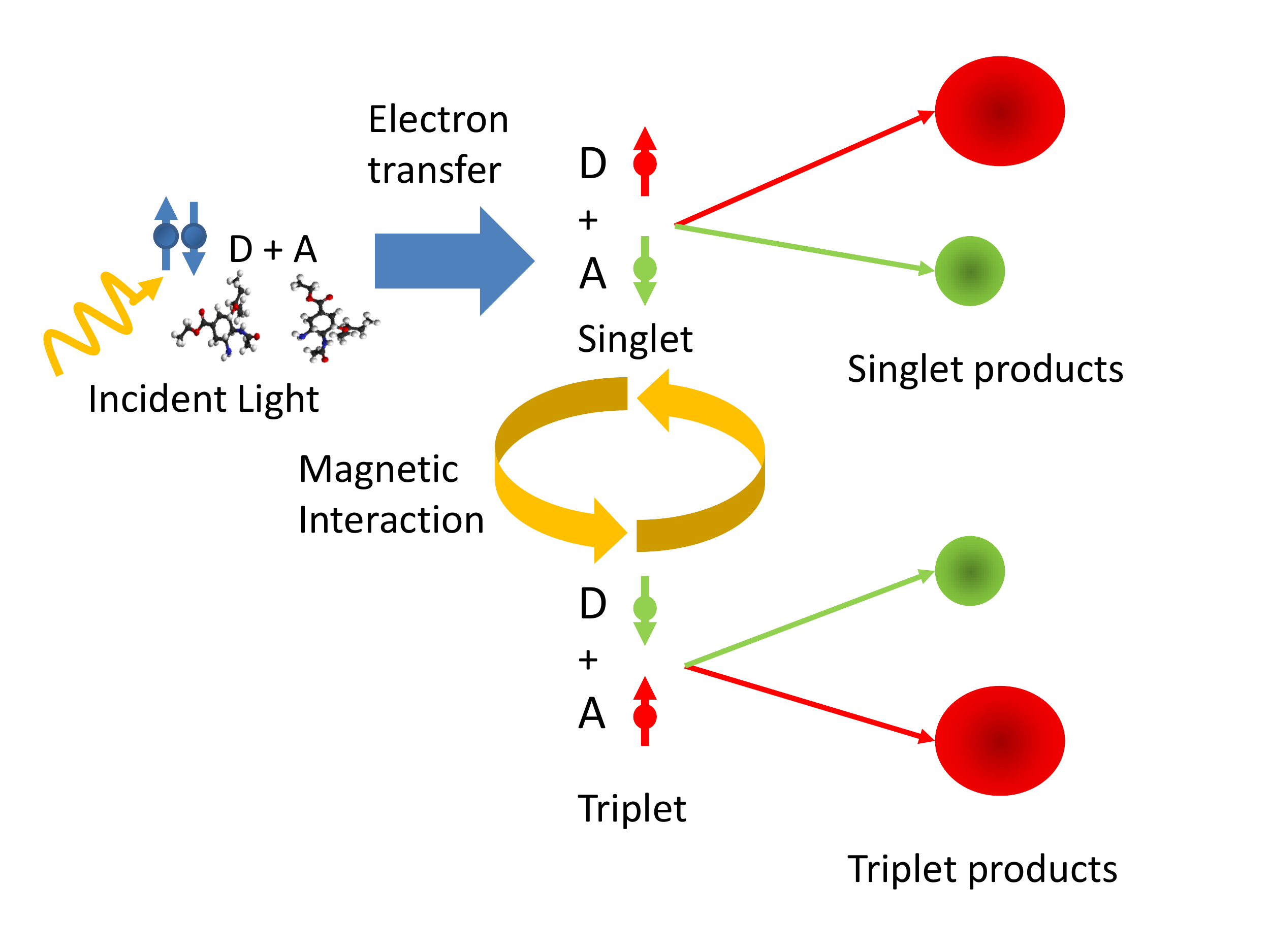}}
\caption{A schematic description of the chemical compass of avian magneto-reception on the basis of the radical-pair mechanism. A photon excites two electrons from the ground state to form a radical pair distributed across two molecules. The local hyperfine interaction leads to transitions between singlet and triplet space whose rate is affected by the orientation of the external magnetic field. Reaction products will depend on the the electrons being in the singlet or triplet space.}
\label{RadicalPair}
\end{figure}

Here we briefly outline the idea of the chemical compass based on the radical pair mechanism and stress that it represents an example of the crucial role of the interplay between electronic spin quantum dynamics and the nuclear spin environment of that electron spin (see Fig. \ref{RadicalPair} for a schematic representation of the following). A donor-acceptor pair is initially in its electronic ground state characterized by a paired electron in a singlet state. Absorption of a photon induces an electron transfer of a single electron from the donor to the acceptor thus creating a radical pair, that is, two molecules with an unpaired electron each. For simplicity we assume that the electronic spin state remains unaffected in this step so that the electrons remain in a singlet state. At this stage no magnetic field sensitivity can be expected as the spin singlet state is rotationally symmetric and hence insensitive to the orientation and magnitude of the external magnetic field.
Hence, a further ingredient is required to break this symmetry. This ingredient is provided by the nuclear spin environment of the donor and acceptor molecules. Due to the distance dependence of the dipolar interaction between electron and nuclear spin the unpaired electrons on donor and acceptor see dominantly uncorrelated interactions which induce symmetry-breaking transitions from the singlet to the triplet manifold. The Hamiltonian of the radical-pair plus nuclear environment is given by
\begin{equation}
    H = \sum_{i=D,A} \sum_j {\bf s}_i {\bf T}_{ij} {\bf I}_{ij} - g \mu_B {\bf B}({\bf s}_A + {\bf s}_D)
\end{equation}
where ${\bf I}_{ij}$ and ${\bf s}_i$ are the nuclear and electron spin operators respectively while ${\bf T}_{ij}$ denotes the hyperfine coupling tensor. ${\bf B}$ denotes the external magnetic field, $\mu_0$ the Bohr magneton and $g$ the gyromagnetic ratio. These transitions and the resulting time varying population of the singlet and the triplet manifold will now be dependent of the relative orientation of the molecules with respect to the external magnetic field via the anisotropy of the hyperfine interaction as well as the strength of the magnetic field as it dictates the energy splitting in the triplet manifold.

Relative population differences between singlet and triplet manifolds result in different rates of occurrence of chemical products. A donor acceptor pair in the singlet state may either recombine to reach the groundstate or lead to a reaction product at rate $k_S$. Spins in the triplet state cannot recombine to reach the ground state but may lead to different reaction products at rate $k_T$. Hence the amount and nature of reaction products will depend on the relative populations of the singlet and triplet manifolds whose time evolution is usually described by a simple phenomenological quantum master equation (the Haberkorn approach \cite{SteinerU89})
\begin{equation}
    \frac{d}{dt}\rho = -\frac{i}{\hbar}[H,\rho] - k_S(Q_S\rho + \rho Q_S) - k_T(Q_T\rho + \rho Q_T)
\end{equation}
where $Q_S$ ($Q_T$) are the projectors onto the singlet (triplet) subspace. The singlet yield is then determined as $k_S \int tr[\rho(t)Q_S] dt$.

The fact that local interactions between electron spins and their nuclear environment lead to a breaking of the symmetry of the electronic spin state gives a first indication of the importance of the presence of local environments. It does not however reveal the full depth of the role of the system-environment interaction. Indeed, contrary to the assumption of early studies, it is not merely the coherence properties of the electron spin state that determine the sensitivity of the chemical compass. To gain a full understanding of the chemical compass and the role of coherence one must study the entire system composed of electron spins and nuclear spins \cite{CaiP13}. It turns out that it is the coherence of the initial state (electronic singlet state and unpolarized nuclear spins) in the basis made up of the eigenstates of the full system in the absence of a magnetic field that has predictive power of the sensitivity of the chemical compass. The principal reason is down to the fact that the coherences in system and environment that are present initially will accumulate phase factors due to the presence of an external magnetic field which in turn will be seen, when observing the electron spin only, as oscillations between singlet and triplet states.
Indeed, a large amount of initial coherence in this picture correlates strongly with high sensitivity of the chemical compass and the impact of different types of decoherence can be predicted from it. Hence it is only when studying the properties of system and environment that we can gain and understanding of the chemical compass and an opportunity to predict optimal designs \cite{CaiP13}.

This example brings out quite clearly that it is the cooperation of electron spin quantum dynamics on the one hand and the interaction with the nuclear spin environment of these electrons on the other hand that lead to the magnetic field sensitivity of the radical pair mechanism. Without the environment no magnetic field sensitivity can be expected. Therefore a more refined picture emerges when the structure of the nuclear spin environment is taken account of by including it in the system dynamics. From this viewpoint the chemical compass can be seen as an environment assisted quantum interferometer thus affirming avian magneto-reception as an example of quantum biology that benefits from an interplay of system-environment interaction \cite{CaiP13}.

\subsection{Structure adaption}
In the preceding section we have presented mechanisms by which the quantum dynamics of electronic systems can enter a beneficial interplay with the dynamics originating from its interaction with an environment. The principles expounded here are more general though and apply to any quantum network whose task may include for example transport or sensing and that is in contact with an environment of charges, spins or vibrations. We have also explained that it is likely that nature has optimized both the properties of its quantum networks {\em and} the structure of the environment to achieve optimal performance.

Here we would like to summarize briefly some of the means that nature has available for achieving this structural adaptation and tuning for optimality. There are two principal aspects that can be controlled, the electronic network and its environment. While in many respects it appears easier to achieve considerable changes in the electronic structure as compared to the vibrational environment it is likely that both will have been subject to evolutionary adaptation.

{\em Position \& Orientation: Interaction strength --} The protein can arrange molecules, such as chromophores, controlling their respective distances as well as the relative orientation of their dipole-moments. Thanks to the $1/r^3$ distance dependence of the dipolar interaction and its angular dependence $3({\bf D}_{i}\cdot{\hat r}_{ij})({\bf D}_{j}\cdot{\hat r}_{ij})-({\bf D}_{i}\cdot{\bf D}_{j})$, where ${\bf D}_{i}$ is the optical dipole moment and ${\hat r}_{ij}$ the unit vector connecting the two molecules, significant changes of the interaction strength and even its sign can be achieved.

{\em Static properties of Local Environments: Onsite excitation energies --} In vacuum two realizations of the same molecule, such as chlorophyll are identical in all its aspects. In a protein scaffold however their properties can vary considerably. These variations are being controlled by the local environment such as the presence and distance of partial charges or even free charges that will affect the local excitation energy of a molecule through their electrostatic interaction. These changes can be significant and can reach the range of $100's$ of wavenumbers or more even for very small distance changes in the environment. For larger structural changes of the protein scaffold, e.g. due to variations in pH-value or other more drastic changes the shift in wavelength can even reach dozens of nanometers. Thus the arrangement of the local environment is of crucial importance for the definition of the electronic properties of a biological quantum network.

{\em Dynamics of Local Environments: Dephasing noise --} The structure of the local environment plays another crucial role as the motion of partial charges or even free charges will lead to fluctuations in the excitation energy of the nearby sites. It is this process for example that couples the vibrational motion of proteins and molecules to their electronic energies and is thus the cause of both dephasing and the interaction of long-lived vibrational modes with electronic degrees of freedom.

{\em Vibrational and spin environments --} The spectral density describing the interaction of electronic degrees of freedom and their environment is a combination of two aspects, both of which may be tuned. Firstly, the coupling strength of individual environmental modes to the system which in turn depends on the amplitude of the motion of charges due to the vibration relative to the site of interest and the magnitude and number of these charges. Secondly, the density of vibrational modes which is determined by the quasi-continuous mode density of the protein as well as the discrete modes provided by the specific molecules that are realizing the quantum network. For spin environments, e.g. the nuclear spin environment in magneto-reception, the structure of the relevant molecule and the form of the electronic wave-function determines the strength and anisotropy of the hyperfine interaction.

{\em PC card principle --} Besides the adjustment of the broad vibrational spectrum of the protein a key principle is the ability to affect the discrete part of the vibrational spectra by the choice of the molecule that realizes the quantum network. Each molecule will be characterized by a set of vibrational modes of varying lifetime. Depending on the fit of the molecule in the protein scaffold the lifetime and coupling of these modes and thus their contribution to the overall spectral density will be affected. In this way, it is conceivable that similar to a PC card the insertion of a specific molecule may allow for certain functionality to be achieved. Likewise the phonon assisted tunneling process in olfaction suggests that this process may be used to recognise molecules and to initiate charge transfer processes.

{\em Enhancing or decreasing noise --} The presence of environmental noise may or may not be of advantage for a biological system. Certainly, strong featureless noise can affect the ability to act and react specifically to external input. Hence, in various circumstances it may be essential to reduce the level of noise for example by protecting a molecule inside a protein structure or a hydrophobic binding pocket. On the other hand it may on occasion be beneficial to increase the level of noise by arranging easily movable charges close to a site which will respond strongly to vibrations and thus create significant changes in the excitation energy of the site.

These are some of the principal means by which biological systems may affect both their quantum dynamical networks as well as the environment around them to achieve optimal performance. This suggests that indeed nature has at its hand a wide variety of possibilities for tuning quantum networks and their environments to achieve robust and efficient devices.

\section{Numerical description of quantum dynamics in structured environments}
One of the key messages of the preceding sections is the special role of the interplay between the quantum dynamics of a system and its environment, in particular when this environment does not merely represent white noise but possesses structure. Furthermore, it has become clear that the optimal operating regime for quantum bio dynamics tends to favour parameter ranges in which the interaction between system components is comparable to the interaction of these system components and their environment.

Both features are not well modeled by the traditional perturbative treatments that lead to master equations of Lindblad \cite{Lindblad76,GoriniKS76}, Redfield \cite{Redfield57} or modified
Redfield type \cite{ZhangMC+98,RengerM03,AdolphsR06} (see \cite{RivasH12} for a review of master equations). Indeed the essential importance of the non-Markovian nature of the system-environment interaction calls for the development of non-perturbative methods that can accurately, certifiably and efficiently model the resulting dynamics. In recent years a wide variety of methods has been developed with the aim of addressing some or all of these issues. In the following we will present the common theoretical setting in which these methods are formulated and then briefly outline key aspects of these methods. Details can be intricate and will be left for further reading.

\subsubsection{Standard model of open-quantum system}
We begin by defining more formally a standard model of an open quantum system as it is relevant to many of the biological settings that we have discussed so far. In this model the quantum system interacts with a macroscopic number of environmental degrees of freedom and the total system and environment state evolves under a purely unitary dynamics. This model is of considerable importance for the description of a wide variety of biological environments. Dissipation and decoherence appear when the system is observed without any knowledge of the state of its environment, leading to a non-unitary \emph{effective} dynamics for the sub-system's reduced density matrix. The total Hamiltonian of system and environment can be written as
\begin{equation}
    H = H_{S} + H_{E} + H_{I},
\end{equation}
where $H_{S}$ is the Hamiltonian that refers to the degrees of freedom of the open quantum system of interest, $H_{E}$ is the Hamiltonian describing the free evolution of the environment while $H_{I}$ describes the interaction between the open quantum system and its environment.

The environment is assumed to be well described by a continuum of harmonic oscillator degrees of freedom labeled by some real number. The internal dynamics of these harmonic oscillators is described by bosonic modes and is thus given by the Hamiltonian
\begin{equation}
    H_{E}=\int_{0}^{x_{\mathrm{max}}}dx\; g(x) a_{x}^{\dagger}a_{x}.
\end{equation}
In a physical context $x$ would usually represent some continuous real variable such as the frequency or momentum of each mode while $x_{\mathrm{max}}$ is the maximum value that this variable may assume. The creation and annihilation operators satisfy the continuum bosonic commutation rules $[a_{x},a_{y}^{\dagger}]=\delta(x-y)$ with the Dirac delta function $\delta(x-y)$. We have adopted a continuum description of the environment but discrete environments can be treated in the same way by replacing the integral by a summation.

The internal dynamics of the system is described by a completely general Hamilton operator $H_{S}$. For the interaction between the system and its environment we assume a linear coupling between arbitrary system operators and the position operator of the environment
\begin{equation}
    H_I = \int_{0}^{x_{\mathrm{max}}}dxh(x)\hat{A}(a_{x}+a^{\dagger}_{x}),
\end{equation}
where $\hat{A}$ is an arbitrary operator of the open quantum system and the (real) coupling function $h(x)$ describes the coupling strength with each mode. It is convenient, though not essential, to consider $g(x)=x$ \cite{footnote1}. In that case the total Hamiltonian for system plus environment is given by
\begin{equation}\label{totalHamiltonian}
    H = H_{S}+\int_{0}^{x_{\mathrm{max}}} \!\!\!\! dx\; x a_{x}^{\dagger}a_{x}
    +\int_{0}^{x_{\mathrm{max}}}dxh(x)\hat{A}(a_{x}+a_{x}^{\dagger}).
\end{equation}
For Gaussian initial states of the environment \cite{footnote2}, the dynamics induced in the quantum system $S$ by its interaction with the environment is determined by the spectral density $J(\omega)$ \cite{LeggettCD+87,Weiss01}. For the continuum model of the reservoir that we are considering, i.e. choosing $g(x)=x$, this function is given by,
\begin{equation}\label{Jhg}
    J(\omega)=\pi h^2(\omega).
\end{equation}
The spectral function thus describes the overall strength of the interaction of the system with the reservoir modes of frequency $\omega$.

Despite the relative simplicity of this model its dynamics is not exactly solvable and one has to resort to numerical methods. This is particularly so, when the environmental spectral density does exhibit considerable structure or when the coupling strength between system and environment lies in the  non-perturbative regime. In the following we discuss a variety of methods that have been developed in recent years to treat this situation.

\subsection{Desiderata}\label{desiderata}
There is a wide variety of simulations methods and in order to evaluate them it will be helpful to
first establish some properties that one may wish a method to possess. Such desiderata for the method include (i) that it is efficient in the system size, (ii) it is able to take account of arbitrary spectral densities without losing efficiency or having to redesign the protocol, (iii) it has the ability to deal with both high, intermediate and low temperatures, and (iv) it should be certifiable, that is it possesses a known and controllable error.

\subsubsection{Time adaptive renormalization group methods}
We begin by presenting an approach to the system-environment interaction that preserves the full information about the environment state, is able to treat arbitrary spectral densities and coupling strengths within a single unified framework with known and controllable error and is computationally efficient in the size of the environment.

The key idea of this method which was developed in \cite{PriorCH+10,ChinRH+10,ChinHP11} is simple but effective. It begins by mapping the spin-boson model exactly onto a 1D system thus permitting the deployment of the time-adaptive density matrix renormalization group (t-DMRG) technique to integrate the time evolution of the full system-environment dynamics efficiently.

\begin{figure}[bht]
\centerline{\includegraphics[width=8cm]{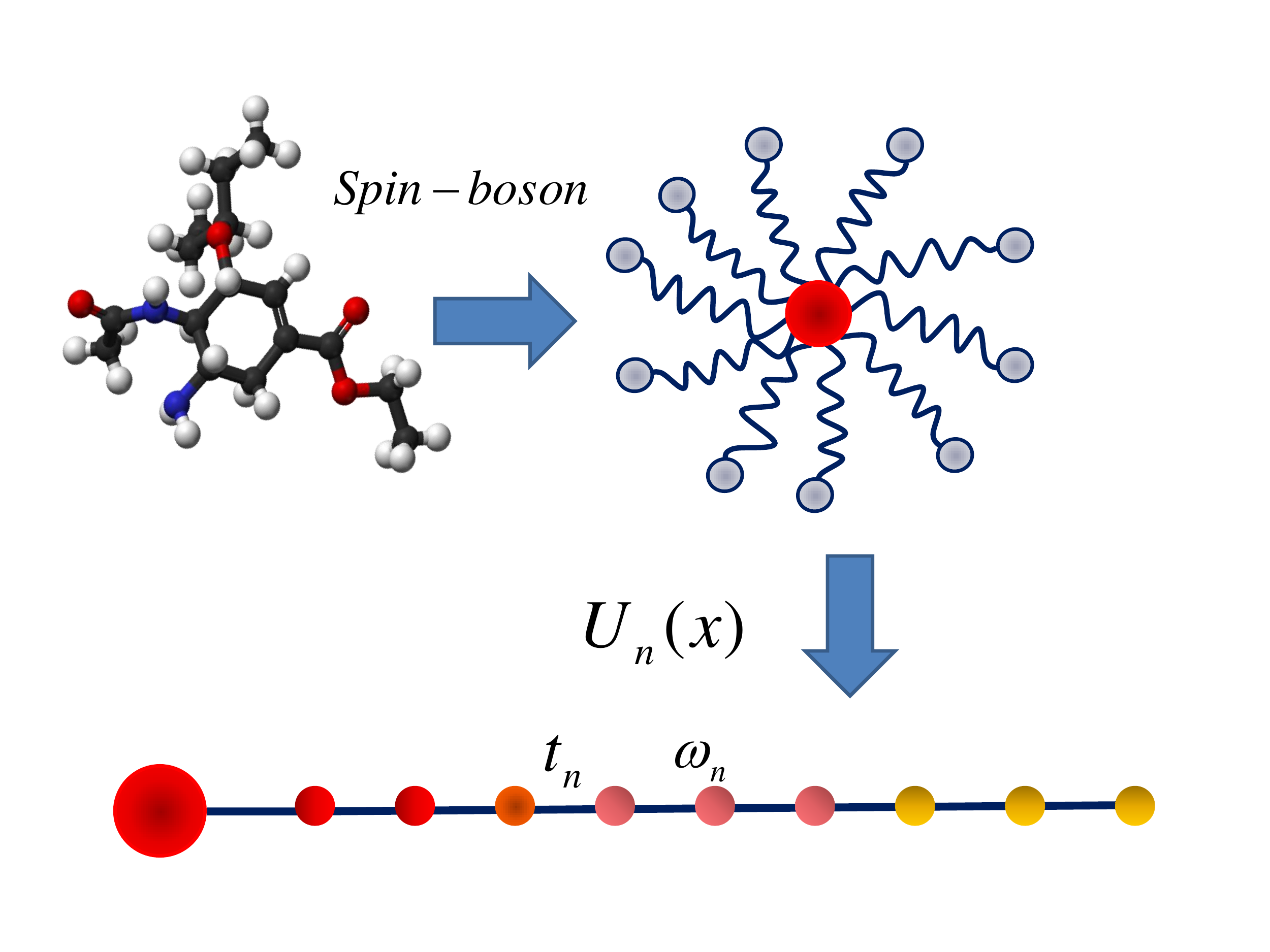}}
\caption{Illustration of the effect of the transformation $U_n(x)$. On the left side a central system is interacting with a reservoir with continuous bosonic or fermionic modes, parametrized by $x$, after $U_n(x)$ the system is the first place of a discrete semi-infinite chain parametrized by $n$.}
\label{FigMapping}
\end{figure}

We begin by demonstrating that a system linearly coupled to a reservoir characterized with a spectral density $J(\omega)$ is unitarily equivalent to a semi-infinite chain with only nearest-neighbors interactions, where the system itself only couples to the first site in the chain (see figure \ref{FigMapping}) \cite{BurkeyC84}. In other words, there exists a unitary operator $U_n(x)$ such that the countably infinite set of new operators
\begin{equation}\label{bn}
    b_{n}^{\dagger}=\int_{0}^{x_{\mathrm{max}}}dx\,U_{n}(x)a_{x}^{\dagger}, \;\;\;
    a_{x}^{\dagger}=\sum_{n}U_{n}(x)b_{n}^{\dagger}
\end{equation}
satisfies the bosonic commutation relations $[b_n,b^\dagger_m]=\delta_{nm}$ and leads to the
transformed Hamiltonian
\[
    H' = H_{S} + c_0\hat{A}(b_{0}+b_{0}^{\dagger}) + \sum_{n=0}^\infty\omega_{n}b_{n}^{\dagger}b_{n}+t_{n}(b_{n+1}^{\dagger}b_{n}+b_{n}^{\dagger}b_{n+1})\label{hc},
\]
where $c_0$, $t_{n}$, $\omega_{n}$ are real constants.
The proof of this statement becomes quite direct after one key observation. Since $J(\omega)$
is positive, we can always define the measure $d\mu(x)=h^2(x)dx$ and write the unitary
\begin{equation}\label{transf}
    U_n(x)=h(x)\tilde{p}_n(x)
\end{equation}
where $\tilde{p}_n(x)$ are some set of real orthonormal polynomials with respect to the
measure $d\mu(x)=h^2(x)dx$ with support on $[0,x_{\mathrm{max}}]$ \cite{BorweinE95}.
Employing $p_0(x)=1$ we immediately find
\begin{eqnarray*}
    H_I &=& \hat{A}(b_{0}+b_{0}^{\dagger}).
\end{eqnarray*}
$H_{E}$ in the new basis is obtained by employing the fact that orthogonal polynomials satisfy the recursion relation $\tilde{p}_{k+1}(x)=(C_kx-A_k)\tilde{p}_k(x)- B_k\tilde{p}_{k-1}(x)$ for $k=0,1,2...$
and $\tilde{p}_{-1}(x)\equiv0$. We find
\begin{eqnarray*}
    H_{E} &=& \sum_{n} [\frac{1}{C_n}b_{n}^{\dagger}b_{n+1}+\frac{A_n}{C_n}b_{n}^{\dagger}b_{n}
    +\frac{B_{n+1}}{C_{n+1}}b_{n+1}^{\dagger}b_{n}]
\end{eqnarray*}
with \cite{ChinRH+10}
\begin{equation*}
    \omega_n=A_n/C_n,\;\;\;\;
    t_n = B_{n+1}/C_{n+1} = 1/C_{n}.
\end{equation*}
This approach provides us with a way to construct an exact mapping, one has just to look for a family of orthogonal polynomials with respect to the measure $d\mu(x)=h^2(x)dx$. This can be done analytically in some important cases (see \cite{ChinRH+10} for examples), however even if the weight $h^2(x)$ is a complicated function, families of orthogonal polynomials can be found by using very stable numerical algorithms such as the ORTHPOL package \cite{Gautschi94}.

\begin{figure}[hbt]
\centerline{\includegraphics[width=8cm]{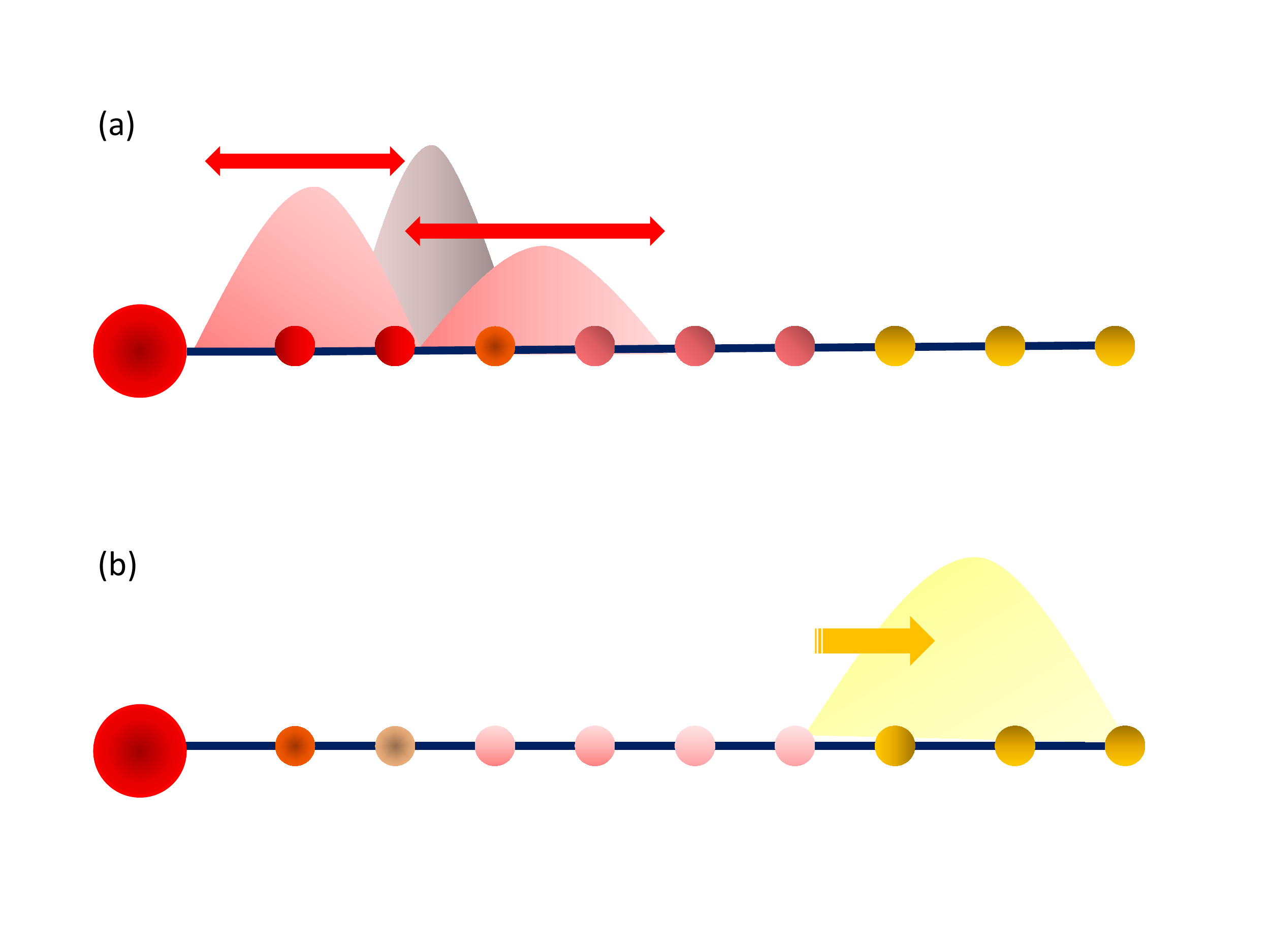}}
\caption{Illustrative sketch of open-system dynamics in the chain representation.
(a) Subsystem initially injects excitations (shown as wave packets) into inhomogenous
region of the chain. Scattering from inhomogeneity causes back action of excitations
on the system at later times and leads to memory effects and non-Markovian
subsystem dynamics. (b) At long times, after multiple scattering, excitations penetrate
into the homogenous region and propagate away from the system without
backscattering. This leads to irreversible and Markovian excitation absorption by the
environment.
}
\label{ChainDynamics}
\end{figure}

The reformulation of the system-environment interaction in the chain picture allows us to apply t-DMRG techniques quite directly. As it can be proven for a wide variety of spectral densities, in the large $n$ limit the chain parameters $\omega_n$ to $t_n$ converge and their ratio approaches $\omega_n/t_n\rightarrow 2$. A translationally invariant harmonic chain with this ratio $\omega_n/t_n$ has a gapless dispersion and excitations can escape down the chain without being scattered back towards the system (see Fig. \ref{ChainDynamics} for illustration). Thus the buildup of excitations in any region of the chain is limited over time, and correlations can be expected to be bounded. This in turn suggests that t-DMRG will provide an accurate description for long times \cite{EisertCP10}. The universal asymptotics of environments revealed by this analysis are discussed in \cite{ChinRH+10}.
Practical experience, though not mathematical proof, suggests that the method itself appears to satisfy all the desiderata listed above. Indeed it has been essential in the accurate numerical modeling dynamics of complex environments which explain the emergence of long-lived quantum coherence in photosynthetic complex \cite{ChinPR+13} as explained earlier. It should be noted however that, in contrast to some other methods, it becomes computationally less efficient with increasing temperature. On the other hand it easily incorporates time-dependent perturbations of the system, e.g. via laser pulses and it does preserve all the information about the environment and is thus capable to following the dynamics after the excitation of complex wave-packets in the environment.

\subsubsection{Hierarchy equations of motion}
Recently, the numerical hierarchy technique \cite{IshizakiF09a,IshizakiF09b,IshizakiT05}, which has a longstanding history \cite{TanimuraHK77,TanimuraHK77a,TanimuraK89}, has received renewed attention in the context of excitation energy transfer across pigment-protein complexes. This approach is non-perturbative and is capable of interpolating, for example, between the Bloch-Redfield and the F{\"o}rster regimes \cite{IshizakiF09a,IshizakiF09b}. It derives a hierarchy of equations in which the reduced density operator of the system couples to auxiliary operators which in principle allow for the simulation of complex environments. The depth of this hierarchy and the structure of its coefficients depend on the correlation time of the bath and its spectral density. This approach appears sufficiently flexible to take account of spatial correlations in the noise as well. For specific choices of the bath spectral density, namely the Brownian harmonic oscillator and/or high temperatures the temporal bath correlations decay exponentially so that the hierarchy can be terminated early with small but uncontrolled error and one obtains a manageable structure of the hierarchy. An estimate suggests that in this case the set of operators in the hierarchy will scale at least proportional $\tau^k$ where $\tau$ is the correlation time and $k$ is the number of sites in the system to be studied (see \cite{IshizakiCS+10} for a more detailed discussion). More complex spectral densities will lead to considerably more demanding evaluations of the coefficients of the hierarchy. A non-exponential decay of the temporal bath correlations also leads one to estimate an exponential growth of the number of operators in the hierarchy. It is possible that specific numerical simulations turn out to be more efficient than those estimates suggest but there is no certificate that provides error bounds. In fact, it is a challenge to determine the errors introduced by the various approximation steps that are involved in the numerical hierarchy technique. Hence one has to test for convergence empirically by increasing the depth of the hierarchy until the result does not change significantly any more. Nevertheless, it should be noted, that the hierarchy method has been applied successfully to a dimer \cite{IshizakiF09a,IshizakiF09b} as well as the seven-site FMO complex \cite{IshizakiF09a,DykstraT12,KreisbeckKR+11} with a Brownian harmonic oscillator spectral density of the environment. Note also recent numerically intensive numerical calculations of 2-D spectra \cite{HeinKK+12} using this method and recent algorithmic improvements \cite{ZhuKR+11}.

\subsubsection{Path integral techniques}
A variety of other methods to study transport processes and in consequence also transport in noisy environments have been developed in condensed matter physics. These methods represent various approaches for finding numerically the formal path integral solution of the time evolution. These include the quasi-adiabatic path integral approach \cite{MakarovM94,MakriM95a,MakriM95b,NalbachT10,ThorwartER+09} and the iterative summation of real-time path integrals \cite{WeissET+08} to name just two. These procedures are expected to give good results in the high temperature limit and hence short correlation time of the environments. With decreasing temperatures and thus increasing correlation time the computational effort grows rapidly. Temperatures not too far below the typical system frequencies appear to be accessible \cite{ThorwartRH00}. For highly structured environments in which for example both narrow and broad features are combined these methods find challenges. In this case, sharply peaked modes can be
added to the system and their damping is treated in the bath \cite{ThorwartPG04} but such an approach will be challenging for the treatment of quantum networks when addition of modes make the network itself too high-dimensional for numerical treatment. Path integral methods have been applied mostly to dimers (see e.g.\cite{ThorwartER+09}) and their scaling to larger systems, just as for the transformation approach to be discussed below, remains to be demonstrated. A comparison of the hierarchy and the path integral methods has been carried out recently for a Brownian harmonic oscillator environment for which both methods are expected to be applicable \cite{NalbachIF+10}.

\subsubsection{Other numerical methods}
Many more methods exist which include methods based on a combination of the polaron transformation and a variational ansatz \cite{McCutcheonN11,McCutcheonN11a,KolliNO11}, schemes based on linearization of the environment dynamics \cite{StockT97,ThossS99,DunkelBC08,HuoC10}, semiclassical methods in which the interaction between the system and its vibrational environment are replaced by mean-field type terms thus obviating the need for tensor product structures \cite{ChinPR+13}, schemes based on time-convolutionless master equations \cite{ShabaniMR+11,BreuerP02}, methods based on quantum state diffusion \cite{RodenEW+09,RitschelRS+11} . For a recent review see of simulation methods see \cite{PachonB12}.

\section{Summary \& Conclusions}
The preceding sections have identified and discussed questions of interest to quantum biology.
In this context the unavoidable presence of partially uncontrolled environments in biology, often perceived as noise, has led us to recognize a theme of fundamental importance that is central to the study of quantum effects in biology -- the interplay between quantum coherent dynamics of a system on the one hand and the interaction of this very system with its environment on the other.

It is this unavoidable lack of isolation that provides the boundary conditions under which natural evolution had to operate and therefore it may, in hindsight, not be surprising that nature has found solutions in which optimal biological quantum dynamics tends to be achieved in a regime where the interaction within the quantum system are of the order of its interaction with the environment so that both contributions do not merely coexist but enter a fruitful interplay.

That this is so, is not an accident. It can be understood from simple and generalizable principles that we have identified in our discussions to explain that too much or too little coherence can be detrimental and that it is in fact natural to expect that there is an intermediate regime that is optimal.

Moreover, we argued, biological systems have available a variety of tools that enable them to achieve these optimal regimes in a controlled fashion. Indeed, by the use of protein structure to arrange molecules and their local environment they are capable of tuning the properties of transport or sensory networks and, at the same time, adjust the environment of these networks by providing isolation or for example by inserting specific molecules with desirable vibrational or spin properties to fashion an environment. It is this toolbox that allows for mutual tuning through evolutionary adaptation which can then be used to achieve optimal performance whose origin we understand from generalizable design principles.

The importance of these design principles goes beyond merely understanding what has been created already but also paves the way by which these principles, when spelled out and made quantitative, can allow for the rational design of optimal structures as well as the execution of optimized experiments by which to amplify and verify quantum effects in biology.

We have followed this path in the preceding sections and have determined and discussed such design principles and have then applied them to bring under one umbrella the phenomena of photosynthesis, olfaction and avian magneto-reception. We hope that guided by the considerations and principles presented here we will be able to add to photosynthesis, avian magneto-reception and olfaction, the three clouds at the otherwise blue sky of classical biology, and discover many more biological phenomena for which quantum effects are of fundamental importance and thus come to be seen as the seeds from which a rich phenomenology of quantum effects in biology may grow.

\section{Acknowledgements} Over the course of the last few years we have benefitted from conversations and collaboration on topics of quantum biology with a wide variety of researchers, in particular J. Almeida, A. Aspuru-Guzik, A. Bayat, J.M. Cai, T. Calarco, F. Caruso, F. Caycedo-Soler, A.~W. Chin, A. Datta, M. del Rey, G.~S. Engel, F. Jelezko, R. Ghosh, S. Montangero, A. Olaya-Castro, H. Plenio, J. Prior, Th. Renger, E. Romero, E. Solano, L. Turin, R. van Grondelle and A. Vaziri. We gratefully acknowledge support by the EU STREP project PAPETS, the ERC Synergy grant BioQ and an Alexander von Humboldt Professorship.


\begin{thebibliography}{200}
\bibitem{Footnote0} The environmental spectral density is a combination of mode density in the environment and the strength of interaction between the system and each of these modes \cite{LeggettCD+87}.
\bibitem{footnote1} In fact a redefinition of modes can always be made to achieve this choice with a suitable redefinition of $h(x)$.
\bibitem{footnote2} These include thermal states and states displaced in phase space which covers a wide class of reasonable initial states in biological systems. It will be highly challenging to create states that differ significantly from these.
\end{thebibliography}
\end{document}